\documentclass{article}

\usepackage{arxiv}

\usepackage[utf8]{inputenc} % allow utf-8 input
\usepackage[T1]{fontenc}    % use 8-bit T1 fonts
\usepackage{hyperref}       % hyperlinks
\usepackage{url}            % simple URL typesetting
\usepackage{booktabs}       % professional-quality tables
\usepackage{amsfonts}       % blackboard math symbols
\usepackage{nicefrac}       % compact symbols for 1/2, etc.
\usepackage{microtype}      % microtypography
\usepackage{amsmath}
\usepackage{cleveref}       % smart cross-referencing
\usepackage{lipsum}         % Can be removed after putting your text content
\usepackage{graphicx}
\usepackage{doi}
\usepackage{ltablex}

\title{Economic Complexity in Mono-Partite Networks}

\usepackage{authblk}

\newbox{\orcid}\sbox{\orcid}{\includegraphics[scale=0.06]{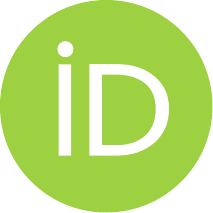}} 

\author[1]{%
	\href{https://orcid.org/0000-0003-3221-5973}{\usebox{\orcid}\hspace{1mm}Vito D.~P.~Servedio\thanks{\texttt{v.servedio@gmail.com}}}%
}
\author[2,3,4]{%
	\href{https://orcid.org/0009-0004-3971-8025}{\usebox{\orcid}\hspace{1mm}Alessandro Bellina}%
}
\author[5]{%
	\href{https://orcid.org/0009-0007-5545-907X}{\usebox{\orcid}\hspace{1mm}Emanuele Cal\`o}%
}
\author[1,2,4,6]{%
	\href{https://orcid.org/0000-0002-3127-5336}{\usebox{\orcid}\hspace{1mm}Giordano De Marzo}%
}
\affil[1]{Complexity Science Hub, Josefstaedter Strasse 39, 1080, Vienna, Austria}
\affil[2]{Centro Ricerche Enrico Fermi, Piazza del Viminale, 1, I-00184 Rome, Italy}
\affil[3]{Sony Computer Science Laboratories - Rome, Joint Initiative CREF-SONY, Centro Ricerche Enrico Fermi, Via Panisperna 89/A, 00184, Rome, Italy}
\affil[4]{Sapienza University of Rome, Physics Dept., P.le A. Moro, 2, I-00185 Rome, Italy}
\affil[5]{IMT School for Advanced Studies Lucca, P.zza San Francesco 19, 55100 Lucca, Italy}
\affil[6]{University of Konstanz, Universitaetstrasse 10, 78457 Konstanz, Germany}
\renewcommand{\shorttitle}{Economic Complexity in Mono-Partite Networks}

\hypersetup{
pdftitle={Economic Complexity in Mono-Partite Networks},
pdfsubject={physics.soc-ph, econ.GN},
pdfauthor={Vito D.~P.~Servedio, Alessandro Bellina, Emanuele Cal\`o, Giordano De Marzo},
pdfkeywords={Complex Networks; Economic Complexity Index; Economic Fitness Complexity},
}

\newcommand{\Eq}[1]{Eq.~(#1)}

\begin{document}
\maketitle

\begin{abstract}
Initially designed to predict and explain the economic trajectories of countries, cities, and regions, economic complexity has been found applicable in diverse contexts such as ecology and chess openings. 
The success of economic complexity stems from its capacity to assess hidden capabilities within a system indirectly.
The existing algorithms for economic complexity operate only when the underlying interaction topology conforms to a bipartite graph. 
A single link disrupting the bipartite structure renders these algorithms inapplicable, even if the weight of that link is tiny compared to others.
This paper presents a novel extension of economic complexity to encompass any graph, overcoming the constraints of bipartite structures.
Additionally, it introduces fitness centrality and orthofitness centrality as new centrality measures in graphs. 
Fitness Centrality emerges as a promising metric for assessing node vulnerability, akin to node betweenness centrality. 
Furthermore, we unveil the cost functions that drive the minimization procedures underlying the economic complexity index and fitness centrality algorithms. 
This extension broadens the scope of economic complexity analysis, enabling its application in diverse network structures beyond bipartite graphs.
\end{abstract}

% keywords can be removed

\section{Introduction}
Network-based algorithms, popularized by Google's PageRank algorithm \cite{brin1998anatomy}, have revolutionized our approach to analyzing and ranking websites and, more generally, networks. Traditional methods of ranking by content analysis — whether manual or automatic — are impractical for large networks and inherently arbitrary, relying on subjective weight assignments to website features. In contrast, algorithms like PageRank and Kleinberg's authority and hub concept \cite{kleinberg1999authoritative} employ a fundamentally different paradigm. By leveraging the interconnected nature of the World Wide Web, these algorithms infer the quality and importance of a website based on its relationships with others, effectively utilizing a graph-based structure for a more objective and scalable ranking system. In simple terms, the idea is that the quality of a website can be estimated as the number of other websites pointing to it, weighted by the quality of these websites themselves. This translates into the linear iterative map underlying the PageRank algorithm. This shift towards using the web's inherent network properties has provided a scalable website ranking solution and introduced a new era in data analysis. 

Following Google's success, several network-based algorithms have emerged or gained popularity, enabling the effective analysis of the vast datasets of the Internet era. Even as deep learning achieves unparalleled success, these algorithms remain among the most interpretable and, in specific contexts, continue to outperform more conventional machine learning approaches \cite{albora2023sapling}. Notably, the PageRank algorithm and its derivatives have found applications across a broad spectrum of systems — from food webs to financial networks and social media \cite{allesina2009googling, battiston2012debtrank, weng2010twitterrank}. These applications have facilitated the identification of key species within ecosystems, pinpointed vulnerable financial institutions, and recognized influential individuals on social networks, underscoring the algorithms' adaptability and impact across diverse domains.

Despite PageRank's versatility, its application is less straightforward in bipartite networks, characterized by two distinct classes of nodes linked across but not within classes.  For example, a streaming platform's user-content interaction can be modeled as a bipartite graph, with edges connecting users to the content they consume. One of the most relevant and influential network theory applications is recommendation algorithms. In particular, collaborative-filtering recommendation algorithms \cite{resnick1994grouplens, linden2003amazon} leverage this network structure, ignoring content features to focus on network patterns for recommendations. Recognizing the unique challenge that bipartite networks present, another aspect that has gathered particular attention is the identification of algorithms to rank nodes in bipartite networks. Such algorithms must consider the different nature of nodes in the two classes, and therefore, techniques devised for monopartite networks are not very suited. The two most notable approaches, both developed in the context of economic complexity, are the Economic Complexity Index (ECI) \cite{hidalgo2009building} and the Economic Fitness and Complexity algorithm (EFC) \cite{cristelli2013measuring, tacchella2012new}.  

Both ECI and EFC have been introduced to analyze the country-product bipartite network, where a country is linked to the products it exports. The goal is to determine which economies are more industrially advanced and which goods are more sophisticated. ECI shares conceptual similarities with PageRank, relying on linear iterative algorithms that can be likened to a random walk. In contrast, EFC employs a nonlinear approach, which has shown superior capability than ECI in measuring countries' capabilities and product complexity \cite{tacchella2013economic}, eventually allowing to obtain state-of-the-art GDP forecasts \cite{tacchella2018dynamical}. After their introduction, both algorithms have found applications in very diverse systems, such as mutualistic networks, chess communities, urban systems, and many more \cite{munoz2015ranking, demarzo2023quantifying, straccamore2023urban, aufiero2023mapping, hidalgo2021economic}. However, despite their great success in analyzing bipartite networks, to our knowledge, neither ECI of EFC has ever been applied to monopartite networks. This is a very relevant limit since many systems, not necessarily belonging to the economics area, can not be investigated using these algorithms, even if only a few links breaking the bipartiteness are present. For instance, in a food web, most animals may act as prey or predators only, with just a few being prey and predator simultaneously. This results in an almost bipartite network where neither ECI nor EFC can be applied.

The present paper aims to generalize ECI and EFC to monopartite networks, thus allowing us to apply them to a wide range of systems, not just economics. This is done by recasting both algorithms in terms of the adjacency matrix of the network and by exploiting the non-homogeneous version of EFC (NHEFC) \cite{servedio2018new}, which converges even if the matrix is not bipartite. In the following, we will refer to the fitness of nodes in a monopartite network as fitness centrality, a novel approach to studying centrality. We also consider the orthofitness centrality, derived by orthogonalizing the fitness-centrality vector to the vector of node degrees. As mentioned, both algorithms have historically been applied only to bipartite networks, so we first investigate which properties they capture in monopartite ones. In particular, we focus on the well-known Zachary Karate Club network \cite{zachary1977information} and on several toy and synthetic graphs. The analysis shows that ECI acts as a community detection algorithm in monopartite networks. At the same time, fitness centrality provides complementary information to other standard measures, such as PageRank. While high PageRank nodes tend to be connected to many nodes with a high degree, high fitness centrality nodes are connected to many nodes with a low degree. This means that fitness-centrality identifies ``crucial'' nodes, i.e., those with the most nodes that are heavily dependent on them. To better investigate this property, we apply fitness-centrality to identify the ``attack vulnerability'' \cite{barabasi1999emergence} of networks, a term referring to the reduction in network performance resulting from the targeted removal of specific vertices or edges. This issue was extensively explored in \cite{holme2002attack}, wherein various strategies were compared across real-world and artificial networks. Thanks to the early removal of ``crucial'' nodes, fitness-centrality performs exceptionally well in generating the maximal number of disjoint communities and in off-line scenarios, where the whole attack strategy must be computed before initiating the attack. 

In conclusion, the new framework we developed also has several theoretical implications. For instance, we better clarify the equivalence between EFC and the Sinkhorn-Knopp algorithm \cite{mazzilli2024equivalence}, the graph clustering properties of ECI \cite{mealy2019interpreting} and the convergence properties of the non-homogeneous version of EFC \cite{servedio2018new}. All these results bridge the gap between the two most prominent bipartite network algorithms and their applicability to the broad set of monopartite networks, showing that ECI and EFC are not limited to economics and can, instead, prove valuable in various systems.

%%%%%%%%%%%%%%%%%%%%%%%%%%%%%%%%%

\section{Materials and Methods}

\subsection{ECI and EFC}
\noindent
We focus on the two main tools employed for assessing economic complexity. 
The first tool introduced in the literature is the Economic Complexity Index (ECI) \cite{hidalgo2009building}. 
This method facilitates ranking countries and products in the international market by leveraging information from the country-product binary matrix $\mathbf{M}$. 
In this matrix, entries $M_{cp}$ are set to 1 if the country $c$ significantly exports the product $p$, and 0 otherwise, as determined by the Revealed Comparative Advantage \cite{RCA}. 
Here, countries and products represent any bipartite network featuring two classes of nodes $c$ and $p$. 
In fact, we will not apply our analysis to country-product datasets; instead, we will base our examination on a generic network, such as the Zachary Karate Club network \cite{zachary1977information}, described in \ref{app:data}.

The algorithm defining ECI is rooted in the method of reflection. 
Consider $N_c$ as the number of countries and $N_p$ as the number of products. 
The total number of products exported by each country, denoted as $k_c$, and the number of countries exporting a specific product, denoted as $k_p$, are determined by the expressions:
\begin{equation}
    k_c=\sum_{p=1}^{N_p}M_{cp} 
        ~~~\mbox{and}~~~
    k_p=\sum_{c=1}^{N_c}M_{cp}.
\end{equation}
These quantities signify the \textit{diversification} of country $c$ and the \textit{ubiquity} of product $p$, respectively. 
The iterative map used in the algorithm is as follows:
\begin{equation}
    \begin{cases}
    % \displaystyle 
        F^{(n)}_c=\frac{1}{k_c}\sum_{p=1}^{N_p}M_{cp}Q^{(n-1)}_p\\
    % \displaystyle 
        Q^{(n)}_p=\frac{1}{k_p}\sum_{c=1}^{N_c}M_{cp}F^{(n-1)}_c,
    \end{cases}
    \label{eq:eci}
\end{equation}
with initial conditions $F^{(0)}_c = k_c$ and $Q^{(0)}_p = k_p$.
The iterative map of \Eq{\ref{eq:eci}} has to be stopped after a few iterations since it will converge to a constant vector.
The number of iterations is chosen arbitrarily. In the original paper, 18 iterations were used in one section of the paper.
The terms $F^{(n)}_c$ and $Q^{(n)}_p$ and their interplay at different $n$ are respectively referred to as the Economic Complexity Index (ECI) and Product Complexity Index (PCI). 
Sorting countries based on their ECI and products based on their PCI results in their rankings. 
Countries with a high ECI value are generally wealthier and focus on producing products with a high PCI \cite{mealy2019interpreting}. 
These products, in turn, are generally more technologically advanced.
This algorithm, denoted as ECI, captures the interplay between a country's diversification and a product's ubiquity.

The second tool we will consider here is the Economic Fitness Complexity (EFC) \cite{tacchella2012new}.
Like ECI, EFC ranks countries and products based on the country-products binary matrix $\mathbf{M}$. 
The following non-linear map defines EFC:
\begin{equation}
    \begin{cases}
        \tilde{F}^{(n)}_c=\sum_{p=1}^{N_p} M_{cp}Q^{(n-1)}_p\\
        \tilde{Q}^{(n)}_p=\left(\sum_{c=1}^{N_c} M_{cp} / F^{(n-1)}_c\right)^{-1}
    \end{cases}
    %\to
    ~\mbox{with the normalization}~~~
    \begin{cases}
        F^{(n)}_c=\tilde{F}_c^{(n)} / \langle\tilde{F}_c^{(n)}\rangle_c\\
        Q^{(n)}_p=\tilde{Q}_p^{(n)} / \langle\tilde{Q}_p^{(n)}\rangle_p.
    \end{cases}
\label{eq:FcQp}
\end{equation}
The initial conditions are set as $\tilde{F}^{(0)}_c = 1$ for all countries and $\tilde{Q}^{(0)}_p = 1$ for all products.
The idea behind EFC is that a country's fitness is the sum of the complexity of the products it exports (not the average as in ECI) and that the complexity of a product is primarily influenced by countries with the lowest fitness that export it.
If a low-fitness country exports a good, this good has low complexity.
In the EFC algorithm, fitness and complexity values are iteratively updated until convergence.
Due to potential convergence issues in the original EFC algorithm \cite{pugliese2016convergence}, we shall use a regularized version called the Non-Homogeneous-EFC (NHEFC) \cite{servedio2018new}.
In this revised version, the map is expressed as:
\begin{equation}
    \begin{cases}
        {F}^{(n)}_c = \delta^2 + \sum_{p=1}^{N_p} M_{cp}/{P}^{(n-1)}_{p}\\
        {P}^{(n)}_p = 1 + \sum_{c=1}^{N_c} M_{cp}/{F}^{(n-1)}_{c}
    \end{cases}
    ~\mbox{with}~~~
    F^{(0)}_c = P^{(0)}_p = 1 ~~ \forall c,p.
    \label{eq:deltaefc}
\end{equation}
Here, the parameter $\delta$ is chosen to be smaller than the typical value of $M_{cp}$ (i.e., $\delta^2\ll 1$).
In this new metric, ${Q}_p=({P}_p-1)^{-1}$ signifies the complexity of a product, while $P_p$ can be considered as a sort of ``simplicity'' of product $p$. 
Contrary to the original version, the non-homogeneous map is not defined up to a multiplicative constant, eliminating the need for the normalization procedure.
Equation~(\ref{eq:deltaefc}) was obtained from the following symmetric map
\begin{equation}
    \begin{cases}
        {\tilde{F}}^{(n)}_c=\delta+\sum_{p=1}^{N_p}M_{cp}/{\tilde{P}}^{(n-1)}_{p}\\
        {\tilde{P}}^{(n)}_p=\delta+\sum_{c=1}^{N_c}M_{cp}/{\tilde{F}}^{(n-1)}_{c}
    \end{cases}
    \label{eq:symmetricNHEFC}
\end{equation}
after setting $\tilde{F}_c = \delta\cdot F_c$ and $\tilde{P}_p=P_p/\delta$.
Rescaling the fitness and simplicity effectively removes the dependence on the parameter $\delta$ as it tends toward zero \cite{servedio2018new}.
All over this work, we will fix $\delta$ to 0.01 and consider the map converged as soon as the maximum relative change in fitness following one iteration falls below 1\%.

\subsection{ECI matrix representation and spectral analysis}
\noindent
Existing literature indicates that ECI is connected to the spectral properties of specific matrices \cite{caldarelli2012network,mealy2019interpreting}. 
In this section, we will provide a brief overview of the spectral method used for estimating ECI.
We begin with the map from \Eq{\ref{eq:eci}} and write it as
\begin{equation}
    \begin{cases}
        \vec{F}^{(n)}=\mathbf{D}^{-1}\mathbf{M}\;\vec{Q}^{(n-1)}\\
        \vec{Q}^{(n)}=\mathbf{U}^{-1}\mathbf{M}^{T}\;\vec{F}^{(n-1)},
    \end{cases}
    \label{eq:eci_matrix1}
\end{equation}
with the matrices $\mathbf{D}^{-1}$ and $\mathbf{U}^{-1}$ defined as square diagonal matrices, where the diagonal elements are the inverses of country diversification and product ubiquity, respectively. 
These matrices have dimensions $N_c\times N_c$ and $N_p \times N_p$, respectively.
The quantities $\vec{F}^{(n)}$ and $\vec{Q}^{(n)}$ are respectively $N_c$ and $N_p$ dimensional vectors and the superscript $T$ represents matrix transposition.
After a single iteration, a step that separates the evolution of economic complexity from that of product complexity, we get:
\begin{equation}
    \begin{cases}
        \vec{F}^{(n)}=\mathbf{D}^{-1}\mathbf{M}\mathbf{U}^{-1}\mathbf{M}^{T}\;\vec{F}^{(n-2)}\\
        \vec{Q}^{(n)}=\mathbf{U}^{-1}\mathbf{M}^{T}\mathbf{D}^{-1}\mathbf{M}\;\vec{Q}^{(n-2)},
    \end{cases}
    \label{eq:eci_matrix}
\end{equation}
We then define the matrices $\mathbf{S}_1$ and $\mathbf{S}_2$ as $\mathbf{S}_1=\mathbf{M}\mathbf{U}^{-1}\mathbf{M}^{T}$ and $\mathbf{S}_2=\mathbf{M}^T\mathbf{D}^{-1}\mathbf{M}$. 
We notice that these matrices are symmetric, indeed
\begin{equation}
    \begin{cases}
        \mathbf{S}_1^{T} & =\bigl(\mathbf{M}\mathbf{U}^{-1}\mathbf{M}^{T}\bigr)^T=\mathbf{M}\mathbf{U}^{-1}\mathbf{M}^{T}=\mathbf{S}_1\\
        \mathbf{S}_2^{T} & =\bigl(\mathbf{M}^T\mathbf{D}^{-1}\mathbf{M}\bigr)^T=\mathbf{M}^T\mathbf{D}^{-1}\mathbf{M}=\mathbf{S}_2,
    \end{cases}
    \label{eq:eci_ss}
\end{equation}
since $\mathbf{U}^{-1}$ and $\mathbf{D}^{-1}$ are diagonal. 
Finally, defining the matrix $\mathbf{N}_1$ as $\mathbf{N}_1=\mathbf{D}^{-1}\mathbf{S}_1$ and the matrix $\mathbf{N}_2$ as $\mathbf{N}_2=\mathbf{U}^{-1}\mathbf{S}_2$, we can express \Eq{\ref{eq:eci_matrix}} as 
\begin{equation}
    \begin{cases}
        \vec{F}^{(n)}=\mathbf{N}_1\;\vec{F}^{(n-2)}\\
        \vec{Q}^{(n)}=\mathbf{N}_2\;\vec{Q}^{(n-2)}.
    \end{cases}
    \label{eq:eci_normal}
\end{equation}
In terms of the initial conditions, we have
\begin{equation}
    \begin{cases}
        \vec{F}^{(2n)}=\mathbf{N}_1^{n}\;\vec{F}^{(0)}\\
        \vec{Q}^{(2n)}=\mathbf{N}_2^{n}\;\vec{Q}^{(0)}.
    \end{cases}
    \label{eq:eci_normal_zero}
\end{equation}
It is worth noting that if both matrices $\mathbf{U}$ and $\mathbf{D}$ were identity matrices, meaning economic and product complexities were defined as sums instead of averages, then $\mathbf{N}_1$ and $\mathbf{N}_2$ would become Kleinberg matrices. 
These matrices were initially employed to calculate authority and hub scores during the early days of the World Wide Web \cite{kleinberg1999authoritative}, providing an alternative to the PageRank algorithm \cite{brin1998anatomy}.
In our context, both $\mathbf{N}_1$ and $\mathbf{N}_2$ are transition matrices, also called normal matrices. 
This stems from the multiplication of $\mathbf{M}$ and $\mathbf{M}^T$ by $\mathbf{D}^{-1}$ and $\mathbf{U}^{-1}$, respectively, which normalizes their rows, ensuring that the elements in each row add up to one. 
Normal matrices possess real eigenvalues within the range of -1 to 1. 
The principal eigenvector, associated with the eigenvalue one, is a constant non-null vector. 
Consequently, the map defined in \Eq{\ref{eq:eci_normal_zero}} is destined to converge to a vector with constant components, namely the principal vector. 
Therefore, when calculating ECI, the process must be halted after a certain number of iterations to obtain meaningful results. 
This approach is akin to considering a linear combination of the first non-trivial eigenvectors, with the one corresponding to the second largest eigenvalue being the most relevant \cite{caldarelli2012network}.

Considering the matrix $\mathbf{N}_1$ (similar reasoning can be applied to $\mathbf{N}_2)$, its eigenvectors $\vec{\varphi}_i$ with corresponding eigenvalues $\lambda_i$ form a complete (non-orthogonal) basis in the space of $N_c$ dimensional vectors. 
Organizing the eigenvalues $\lambda_i$ in descending order, where $\lambda_1>|\lambda_2|\geq\dots\geq|\lambda_{N_c}|$, we can express $\vec{F}^{(0)}$ as:
\begin{equation}
    \vec{F}^{(0)} =  a_1\vec{\varphi}_1 + a_2\vec{\varphi}_2 + \dots + a_{N_c}\vec{\varphi}_{N_c}.
\end{equation}
Further, rewriting the first equation of \Eq{\ref{eq:eci_normal_zero}} gives:
\begin{equation}
    \begin{split}
        \vec{F}^{(2n)} &= a_1\vec{\varphi}_1 + a_2\lambda_2^n\vec{\varphi}_2 + \dots + a_{N_c}\lambda_{N_c}^n\vec{\varphi}_{N_c}\\
         &= a_1\vec{\varphi}_1 + a_2\lambda_2^n\vec{\varphi}_2 +\lambda_2^n O((\lambda_3/\lambda_2)^n).
    \end{split}
    \label{eq:eci_expansion}
\end{equation}
Consequently, for sufficiently large $n$, the country ranking is determined by the components of $\vec{\varphi}_2$, since $\vec{\varphi}_1$ is constant. 
Hence, computing the eigenvector components associated with the second eigenvalue reveals the country ranking. 
In \ref{app:GS}, we recall the Courant-Fisher-Weil procedure to iteratively determine the second eigenvector of the normal matrix $\mathbf{N}_1$ without solving the entire eigenvalue problem. 
Notably, we cannot choose a constant $\vec{F}^{(0)}$; otherwise, we stay trapped in the principal eigenvector.

%==============================================================
%==============================================================
%==============================================================
\subsection{Mono-partite representation}
\noindent
The information relevant for a bipartite network is encapsulated in the rectangular matrix $\mathbf{M}$ with dimensions $N_c\times N_p$. 
The graph itself is represented by its square adjacency matrix $\mathbf{A}$ with dimensions $(N_c+N_p) \times (N_c+N_p)$. 
For an undirected graph, $\mathbf{A}$ can be expressed as:
\begin{equation}
    \mathbf{A}=
        \begin{pmatrix} \mathbf{0} & \mathbf{M} \\
        \mathbf{M}^T & \mathbf{0}
        \end{pmatrix}.
    \label{eq:A}
\end{equation}
The diagonal blocks of $\mathbf{A}$ consist of zeros --- the first block with dimensions $N_c \times N_c$ and the second with dimensions $N_p \times N_p$. 
Given the bipartite nature of the graph, no links exist between countries or between products.

\noindent
Now, let's express both ECI and NHEFC in terms of the graph's adjacency matrix $\mathbf{A}$.\\[3mm]
\textit{ECI}\\
We introduce the vector $\vec{V}$ with $N_c+N_p$ components, where the first $N_c$ elements represent economic complexity and the last $N_p$ elements represent product complexity. Additionally, we introduce the vector $\vec{k}$, where the first $N_c$ elements denote the diversification of countries and the last $N_p$ components represent the ubiquities of products.
The ECI algorithm can be written as:
\begin{equation}
    V^{(n)}_i = \frac{1}{k_i} \sum_{j=1}^{N_c+N_p} A_{ij} V^{(n-1)}_j ~~\mbox{with}~~~ V^{(0)}_i=k_i.
    \label{eq:eci_map}
\end{equation}
By introducing the diagonal matrix $\mathbf{K}$ with the elements of vector $\vec{k}$ on its diagonal, we can express ECI in matrix form:
\begin{equation}
    \vec{V}^{(n)} = \mathbf{K}^{-1} \mathbf{A}\; \vec{V}^{(n-1)} ~~\mbox{with}~~~ \vec{V}^{(0)}=\vec{k}.
    \label{eq:eci_transition}
\end{equation}
The matrix $\mathbf{K}^{-1} \mathbf{A}$, denoted as $\mathbf{S}$ hereafter, is a normal matrix, as its rows sum to one. 
Such matrices, known as transition matrices in physics, are employed to model random walkers on a network.
In essence, equations like the one presented in \Eq{\ref{eq:eci_transition}} describe a random walker in its n-th step.
The analogy between ECI and a random walk was already pointed out in the original article \cite{hidalgo2007product}.
The reformulation of ECI in terms of the adjacency matrix of the bipartite graph helps us understand better that ECI can be considered a spectral community detection algorithm in graphs \cite{mealy2019interpreting,capocci2005detecting,servedio2005community}.
It might seem that the method of reflections is an unnecessary complication arising from the use of the reduced matrix $\mathbf{M}$ instead of the adjacency matrix $\mathbf{A}$ since all the essential information regarding ECI is already encapsulated in the matrix $\mathbf{S}$.
In practice, there are some advantages to working with the square of $\mathbf{S}$.
This choice is motivated by the fact that the spectrum of $\mathbf{S}$ is symmetric (if $\lambda$ is an eigenvalue, then $-\lambda$ is also), and there are bouncing effects when estimating its eigenvectors recursively. 
Hence, it makes sense to consider the normal matrix $\mathbf{N}_A = \mathbf{S}^2$, which has eigenvalues between zero and one.
This approach effectively splits the problem into two independent random walkers -- one for countries and the other for products -- by creating a random walker moving two steps at a time on a bipartite network. 
Moreover, squaring $\mathbf{S}$ decreases the ratio between the second-largest and third eigenvalues --- $\lambda_3/\lambda_2$ becomes $(\lambda_3/\lambda_2)^2$ --- making the second eigenvector of $\mathbf{N}_A$ already a good approximation of ECI (see \Eq{\ref{eq:eci_expansion}}).\\[3mm]
\textit{NHEFC}\\
Similar to our approach with ECI and PCI, we define a vector $\vec{V}$ with $N_c+N_p$ components. Here, the first $N_c$ components represent countries' fitness, while the remaining $N_p$ components signify the products' simplicity. Adhering to this convention, the symmetric NHEFC map of \Eq{\ref{eq:symmetricNHEFC}} can be expressed as:
\begin{equation}
    V^{(n)}_i = \delta + \sum_{j=1}^{N_c+N_p} A_{ij} / V^{(n-1)}_j ~~\mbox{with}~~~ V_i^{(0)} = 1.
\end{equation}
Alternatively, in a more compact form:
\begin{equation}
    \vec{V}^{(n)} = \delta\cdot\vec{1} + \mathbf{A}\cdot (\vec{V}^{(n-1)})^{-1} ~~\mbox{with}~~~ \vec{V}^{(0)} = \vec{1},
    \label{eq:compact_nhefc}
\end{equation}
where $\vec{1}$ represents the vector with all its components set to one, and $(\vec{V})^{-1}$ is the vector obtained by inverting all the components of $\vec{V}$.\\[3mm]
With the introduction of the adjacency matrix $\mathbf{A}$ in both the ECI and NHEFC formulations, it is straightforward to generalize both algorithms to the case of any graph, where $\mathbf{A}$ no longer needs to be bipartite.
Without bipartiteness in the graph, we no longer differentiate between ECI and PCI, as well as fitness and simplicity. 
Thus, we consider the vector $\vec{V}$ solely as ECI or fitness in both scenarios.

\subsection{Orthofitness}
\label{sec:ortho}
\noindent
One objection regarding the meaningfulness of fitness is its strong correlation with the nodes' degrees, although it is logical that countries exporting more products tend to be fitter. 
To address this, we introduce orthofitness, obtained by orthogonalizing the fitness vector to the vector of node degrees, following the Gram-Schmidt process found in graduate-level textbooks. 
We somehow exploit the analogy of ECI being orthogonal to the diversification vector (see \Eq{\ref{eq:orthoECI}}).
Given a vector $\vec{V}$, we orthogonalize it with respect to another vector $\vec{d}$ using the following process:
\begin{equation}
    \vec{V}_\perp  = \vec{V} - \frac{\vec{V}\cdot\vec{d}}{|d|^2} \vec{d}.
\end{equation}
We denote the resulting vector $\vec{V}_\perp$ as orthofitness.

\subsection{Cost Functions}
\noindent
In most cases, iterative algorithms that converge to a particular solution rely on minimizing a cost function \cite{wright2006numerical}. 
Optimization algorithms \cite{hartmann2002optimization, arora2015optimization} are used to find the best possible solution to a given problem, and they do it by minimizing a particular objective function. 
One example in which optimization algorithms play a crucial role is Machine Learning \cite{zhou2021machine, mahesh2020machine} where the goal is to find a representation of data by minimizing a loss function through gradient descent \cite{ruder2016overview}. 
Linear programming \cite{dantzig2002linear}, spectral decomposition \cite{kannan2009spectral}, and simulated annealing \cite{bertsimas1993simulated} are only a few other examples of these types of algorithms. 

In this section, we will show how both ECI and NHEFC can be considered optimization algorithms because they can be rephrased to solve the problem of minimizing a particular function. 
This objective function is usually defined at the beginning, and then, eventually, iterative equations are found. 
In our case, we must do the inverse process and integrate \Eq{\ref{eq:eci_map}} and \Eq{\ref{eq:compact_nhefc}} to recover the expression of the cost. 
The cost function of ECI is related to the eigenvector findings problem and can be easily interpreted in this sense. 
In the case of NHEFC, the analogy with optimization problems has already been pointed out in \cite{mazzilli2024equivalence} due to the similarity of equations \ref{eq:FcQp} with the Sinkhorn-Knopp algorithm \cite{knight2008sinkhorn} used in optimal transport theory. 
In the following, we will delve into this connection, providing the exact cost function of NHEFC algorithm. 
Details on the derivations of the cost functions for both ECI and NHEFC are reported in \ref{app:cost_derivation}.\\[3mm]
\textit{ECI}\\
By integrating \Eq{\ref{eq:eci_map}}, we can obtain that the solution of the algorithm is equivalent to finding the minimum of the following cost function:
\begin{equation}
    U(\vec{V}) = - \dfrac{1}{2} \sum_{i, j} A_{ij} V_i V_j + \dfrac{1}{2}\sum_i k_i V_i^2
    \label{eq:eci_cost_general}
\end{equation}
This function has the same form of the cost used in standard eigenvector finding algorithms \cite{watkins1993some}. 
It is also formally related to a spin system Hamiltonian, where the first is an interaction term, and the second acts as a regularization. 

The cost function \Eq{\ref{eq:eci_cost_general}} can be also expressed in terms of the Laplacian matrix $L_{ij} = k_i \delta_{ij} - A_{ij}$ of the network $A_{ij}$:
\begin{equation}
    U(\vec{V}) = \frac{1}{4}\sum_{i,j} A_{ij} (V_i - V_j)^2 = \dfrac{1}{2} \sum_{i, j} L_{ij} V_i V_j .
    \label{eq:eci_cost_general_L}
\end{equation}
The minimum of this function is when $V_i = V_j$ for all $i$ and $j$, which is the constant vector. 
This is also the first eigenvector of the Laplacian matrix, representing the solution that ECI algorithm approaches after many iterations. 
On the other hand, the second eigenvector, which represents the ECI index, maximizes the partitioning between the network. 
This is consistent with the fact that ECI is, in the core, a spectral community-based algorithm.\\[3mm]
\textit{NHEFC}\\
In this case, integrating \Eq{\ref{eq:compact_nhefc}} gives:
\begin{equation}
    U(\vec{V}) = \dfrac{1}{2} \sum_{i, j} A_{ij} \dfrac{1}{V_i V_j} - \sum_i \log\dfrac{1}{V_i} + \delta \sum_i \dfrac{1}{V_i}
    \label{eq:efc_cost_general}
\end{equation}
This expression is more complicated than \Eq{\ref{eq:eci_cost_general}}, but still can be interpreted straightforwardly in terms of its minimization. 
In this case, the interaction term contains not the variable itself but its reciprocal.
The result is that any variable assumes a value that accounts for information from the other nodes of the network, working in a similar sense to standard centrality measures. 
However, the way the information is treated is different.

The first regularization is a logarithmic barrier for the interaction term. It also reveals that the natural scale of the fitness variables is the logarithmic one, as already pointed out in several cases (\cite{tacchella2012new,mazzilli2024equivalence}). 
It is important to note that this logarithmic term cannot prevent the divergence of the cost function since the interaction term is hyperbolic. 
Thus, the second regularization from the constant $\delta$ is needed to ensure the algorithm's full convergence (the plain EFC converges in ranks, i.e., when the ranking of countries does not change further). 
Utilizing an analogy with a spin system, $\delta$ behaves like a magnetic field, whose strength can be small, compelling the variables to align eventually.
Equivalently, the factor $\delta$ can be seen as a Lagrange multiplier enforcing the term $\sum_i {V_i}^{-1}$ to be constant.
In both ways, its task is to avoid the divergence of the quantities in the system, connected to small and large $V_i$ components.

%%%%%%%%%%%%%%%%%%%%%%%%%%%%
%%%%%%%%%%%%%%%%%%%%%%%%%%%%
%%%%%%%%%%%%%%%%%%%%%%%%%%%%

%%%%%%%%%%%%%%%%%%%%%%%%%%%%%%%%%%%%%%%%%%

%%%%%%%%%%%%%%%%%%%%%%%%%%%%%%%%%
\section{Results}
\subsection{Fitness as a new centrality measure}
\begin{figure}
    \centering
    \includegraphics[width=0.48\textwidth]{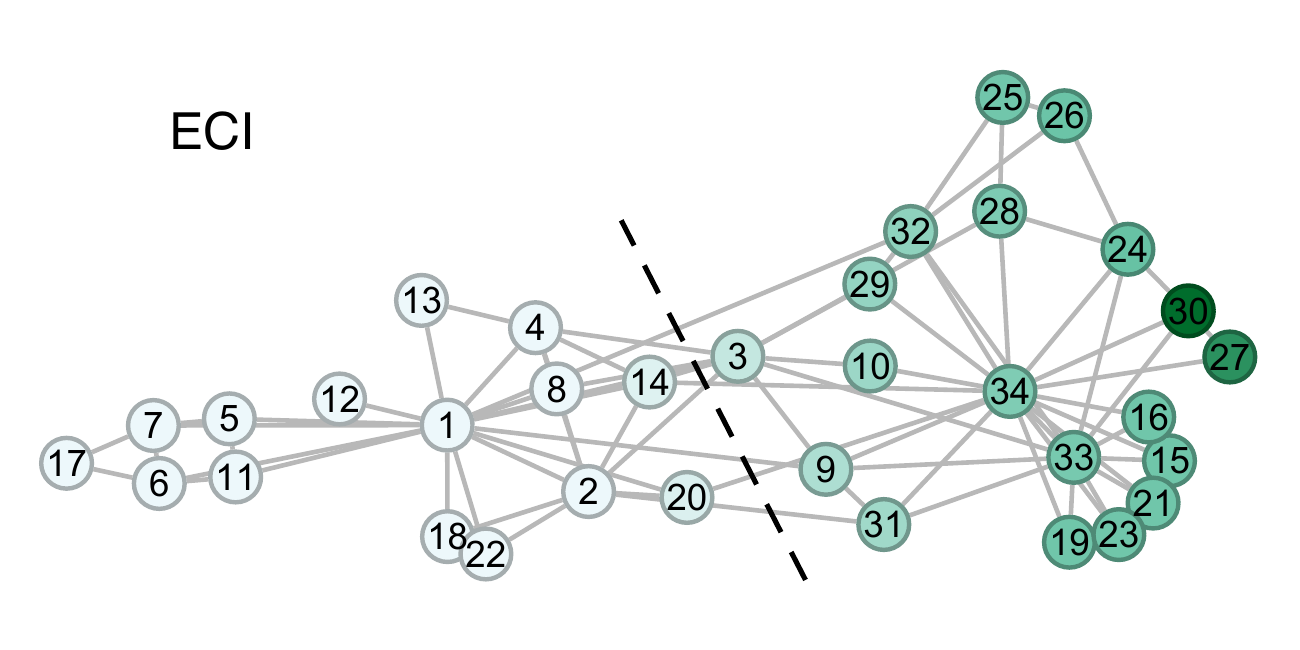}%
    \includegraphics[width=0.48\textwidth]{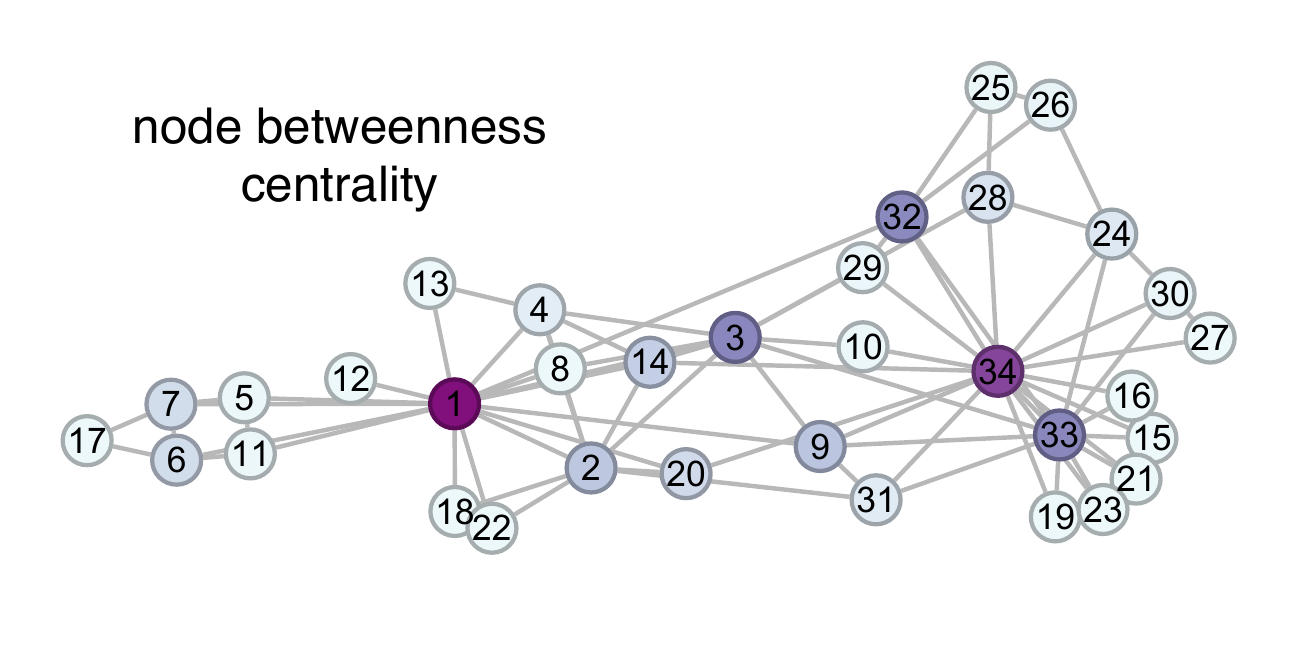}
    \caption{\textbf{ECI and node betweenness centrality estimated on the Zachary Karate Club network.}
        The nodes appear darker when their ECI and node betweenness centrality are higher.
        ECI (left panel) offers the one-dimensional embedding of the graph and identifies its minimum cut.
        Nodes on either side of the dashed line exhibit opposite-sign ECIs.
        The minimum cut almost detects the splitting into two separate communities that took place in reality.
        Only node three is wrongly assigned to the component on the figure's right.
        The node betweenness centrality measures the shortest paths going through nodes. 
        This metric is frequently employed to identify nodes whose removal would result in the most significant disruption to the network.
%        Eigenvector centrality (right panel) prioritizes nodes with higher degrees and their neighboring nodes.
        }
    \label{fig:zkc_eci}
\end{figure}

\noindent
We tested the mono-partite versions of ECI (\Eq{\ref{eq:eci_transition}}) and NHEFC (\Eq{\ref{eq:compact_nhefc}}) on the renowned Zachary Karate Club network, which portrays social interactions within a karate sports club. 
This network has gained prominence as it delineates the club's participants, who split into two groups following two instructors. 
Consequently, it has been extensively utilized to evaluate community detection algorithms in graphs \cite{girvan2002community}, assuming this division is an objective measure of the two primary communities.
The Zachary Karate Club network is undirected and unweighted.
In the left panel of Fig.~\ref{fig:zkc_eci}, we depict the Zachary Karate Club network with nodes' colors coded according to their ECI value.
We observe that ECI effectively identifies the minimum cut of the graph, revealing two distinct communities of nodes. 
Notably, nodes on opposite sides of the dashed line exhibit ECI values of opposite signs. 
Interestingly, the ``force atlas'' method employed to plot the network \cite{gephi} sets the horizontal position of the nodes in agreement with their respective ECI values.
This alignment is not a coincidence, as the force atlas algorithm conceptualizes links within the graph as springs. 
This analogy invokes the Laplacian matrix, whose first non-trivial eigenvectors delineate the graph's embedding, revealing its underlying community structure.
In the right panel of Fig.~\ref{fig:zkc_eci}, we employ a color scheme to represent the nodes based on their betweenness centrality within the network. 
Node betweenness centrality quantifies the shortest paths that traverse a given node. 
This metric is commonly employed to identify nodes with significant traffic flow whose removal could lead to substantial disruptions within the network.

\begin{figure}
    \centering
    \includegraphics[width=0.48\textwidth]{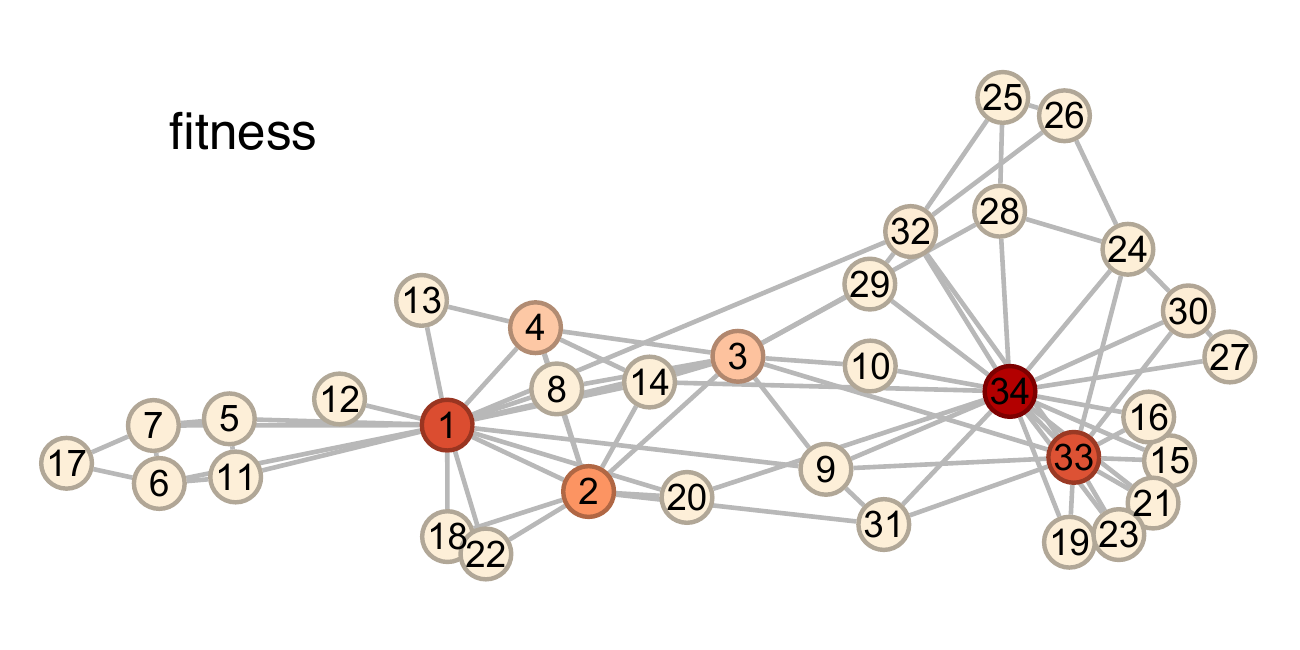}%
    \includegraphics[width=0.48\textwidth]{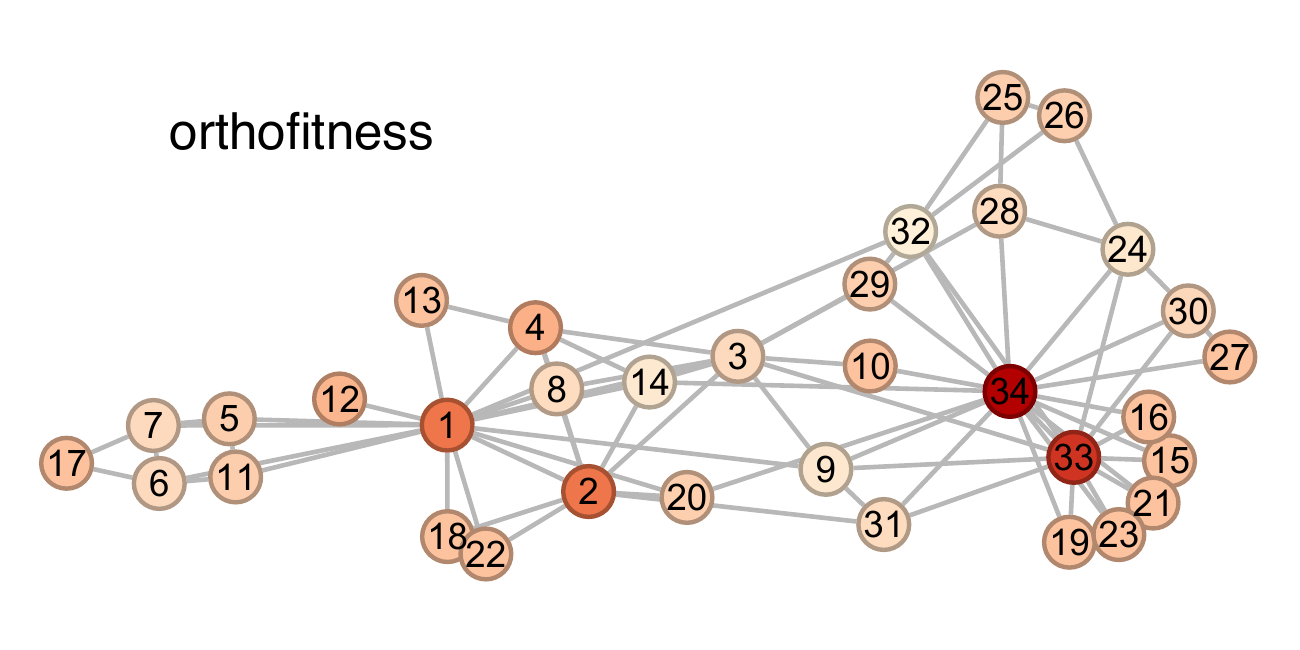}
    \caption{\textbf{Fitness and orthofitness estimated on the Zachary Karate Club network.}
        The nodes are depicted as darker when their fitness (left panel) and orthofitness (right panel) are higher. 
        The orthofitness cancels out the net effect of a high degree and reveals a smoother but similar picture.
        }
    \label{fig:zkc_fitness}
\end{figure}

In Fig.~\ref{fig:zkc_fitness}, we depict the Zachary Karate Club network, with nodes' colors indicating their fitness centrality (left panel) and orthofitness centrality (right panel). 
Notably, in both scenarios, nodes with high degrees are highlighted prominently. 
Both fitness scores tend to assign relatively less significance to the neighbors of high-degree nodes than the eigenvector centrality shown in Fig.~\ref{fig:eigvcentrality}.
When evaluating \Eq{\ref{eq:compact_nhefc}}, we set $\delta$ to 0.01 and determine convergence when the maximum relative change in fitness following one iteration falls below 1\%.

\begin{figure}[t!]
    \centering
    \includegraphics[width=0.9\textwidth]{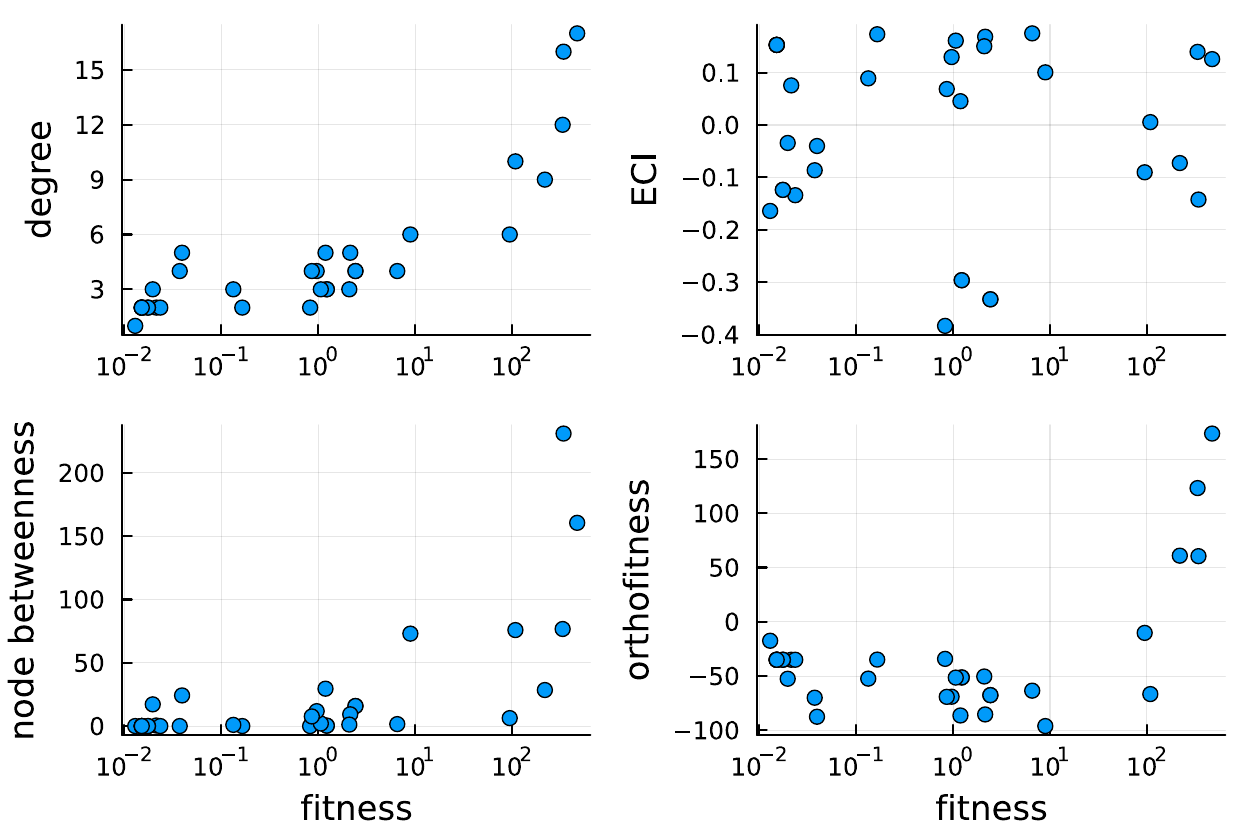}
    \caption{\textbf{Selected centrality measures vs.\ fitness in the Zachary Karate Club network.}
        Each dot in the plots represents a single node within the network. 
        Across all plots, the horizontal axis represents fitness on a logarithmic scale. 
        As expected, we observe a correlation between fitness, high degrees, and high orthofitness.
        The correlation with node betweenness is negligible.
        Precisely quantifying this correlation is beyond the scope of this article. 
        Fitness does not correlate with ECI, which measures the one-dimensional graph embedding -- a conceptually distinct attribute.
                }
    \label{fig:zkc_all_vs_fitness}
\end{figure}

\begin{figure}[t!]
    \centering
    \includegraphics[width=0.9\textwidth]{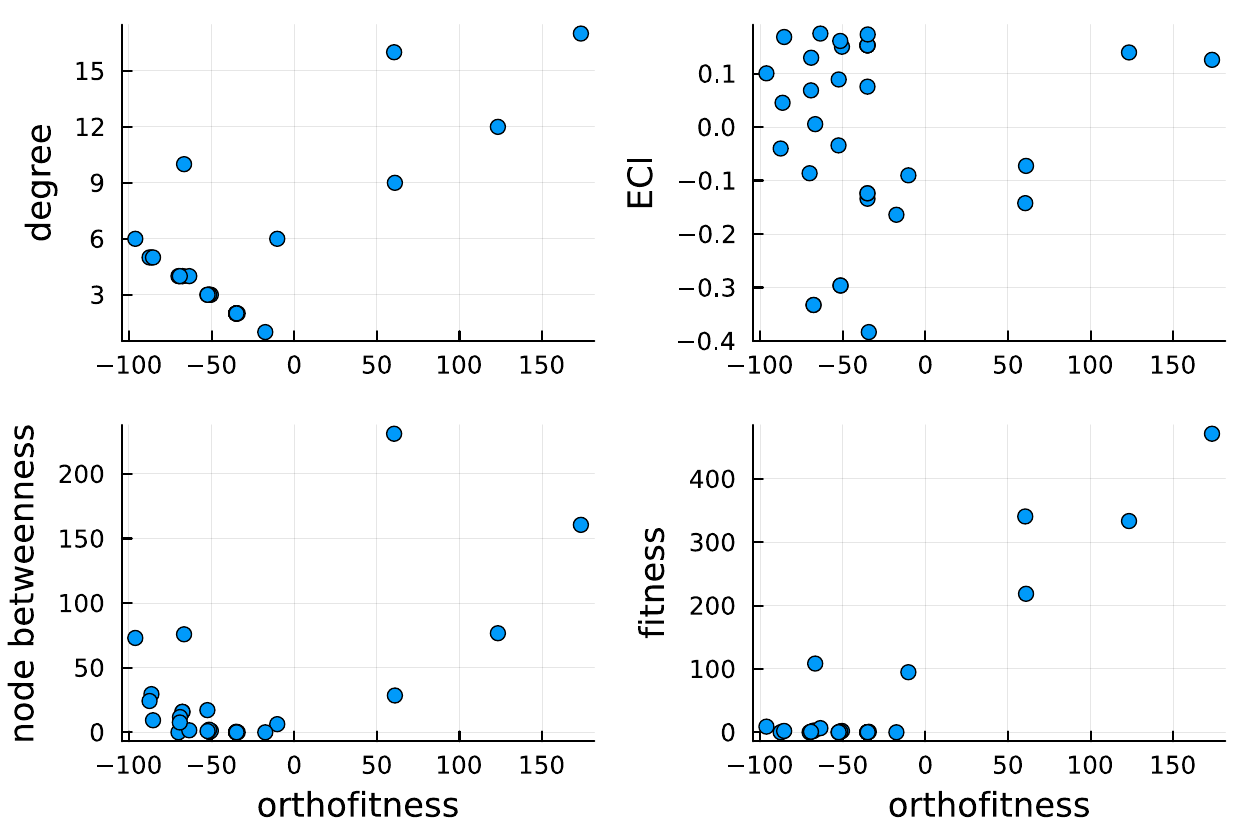}
    \caption{\textbf{Selected centrality measures vs.\ orthofitness in the Zachary Karate Club network.}
        Each dot in the plots represents a single node within the network. 
        Across all plots, the horizontal axis represents orthofitness.
        The same considerations of Fig.~\ref{fig:zkc_all_vs_fitness} for fitness also apply to orthofitness. 
        The bottom right plot is the same as that of Fig.~\ref{fig:zkc_all_vs_fitness} but with both axes on a linear scale.
        }
    \label{fig:zkc_all_vs_orthofitness}
\end{figure}

In Fig.~\ref{fig:zkc_all_vs_fitness} and Fig.~\ref{fig:zkc_all_vs_orthofitness}, we juxtapose the fitness and orthofitness centralities against node degree, ECI, and node betweenness centrality.
Neither fitness centralities correlate with ECI, underscoring their distinctiveness in capturing different graph properties. 
They also show no correlation with node degrees at lower values, but a correlation emerges at higher values.
A mild correlation is found with respect to the betweenness centrality.
Similarly, orthofitness and fitness centralities demonstrate correlation only at higher values.
We employed the node degree as a reliable proxy for PageRank. 
In undirected networks, PageRank closely aligns with node degree \cite{grolmusz2015note}. 
Specifically, it becomes identical when the PageRank algorithm's jumping probability is set to zero. 
However, assigning a non-zero jumping probability in undirected networks is of no practical significance.

\begin{figure}[t!]
    \centering
    \includegraphics[width=1\textwidth]{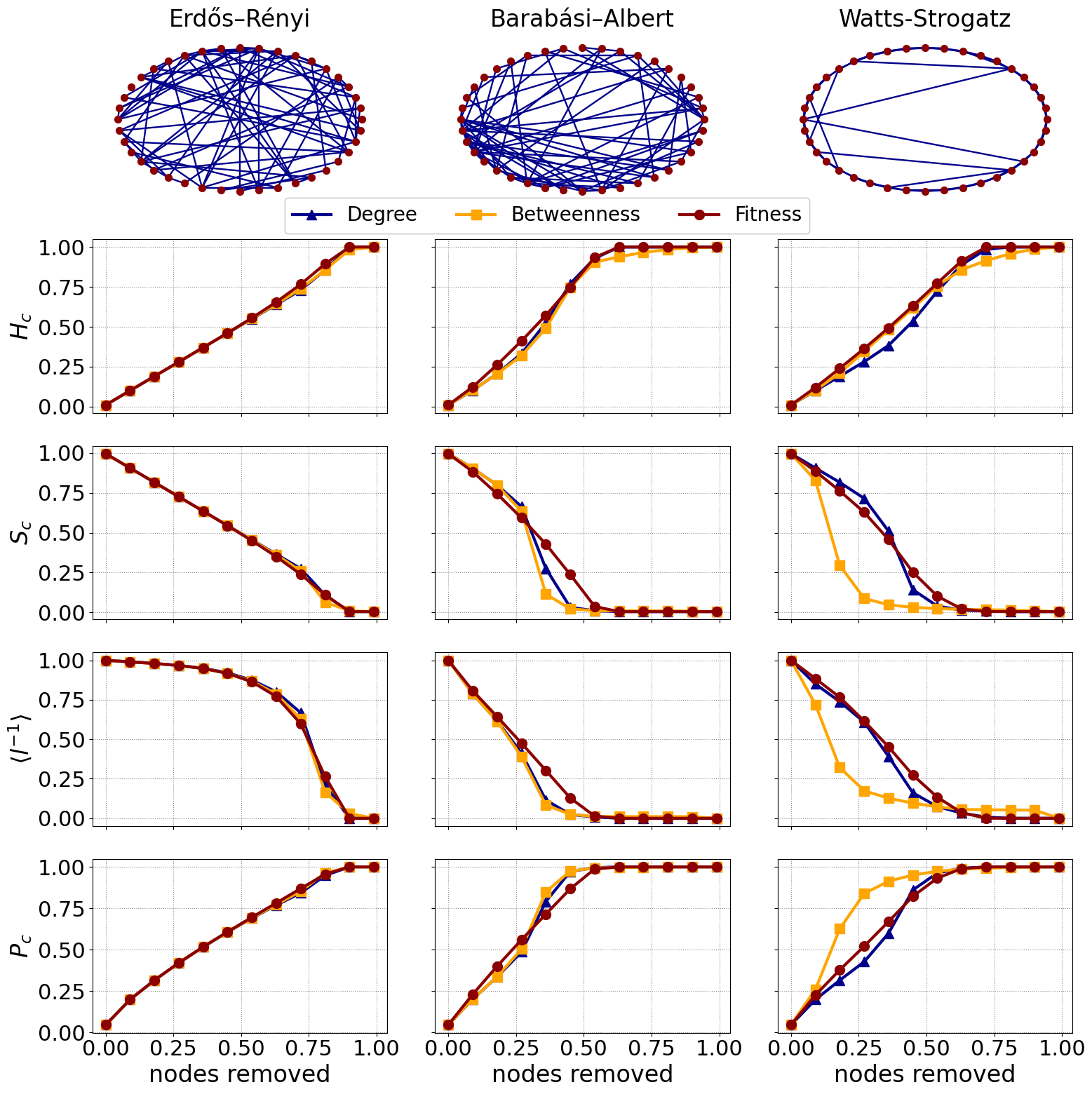} \\
    \caption{\textbf{Attack vulnerability of synthetic networks.}
    Results of the attack vulnerability analysis on three different types of networks. Fitness refers to the original metric of \Eq{\ref{eq:compact_nhefc}}. Each quantity $H_c$, $S_c$, $\langle l^{-1} \rangle$, $P_c$ and the number of nodes removed are normalized to $1$ for visualization. \textbf{First column} Analysis for Erdős–Rényi random networks. Since the degree is Poissonian distributed, all metrics perform equally in $H_c$, $S_c$, $\langle l^{-1} \rangle$, and $P_c$ as expected. \textbf{Second column} Barab\'asi–Albert scale-free network. The fitness strategy is the optimal one for the number of connected components $H_c$. Measuring the size of the largest connected component $S_c$, the fitness performs better than the betweenness until the removal of $\approx 0.3$ of the total nodes, as it also happens for $P_c$. Regarding the inverse geodesic length $\langle l^{-1} \rangle$, the fitness performs as well as the betweenness until the removal of $\approx 0.3$ of the total nodes. \textbf{Third column} Watts-Strogatz small world network. The fitness strategy performs better than the others for $H_c$, while it performs poorly on $S_c$, $\langle l^{-1} \rangle$ and $P_c$. The evaluated performance through the AUC measure is reported in \ref{app:attack}.
                }
    \label{fig:attack_synthetic}
\end{figure}

\subsection{Attack Vulnerability based on fitness score}
\label{sec:attack}
\noindent
In this section, we focus on vertex attacks on networks, analyzing the best techniques that can be used to guide a disruption of such structures. 
The approach involves assigning nodes a specific score, such as degree or a centrality measure like betweenness, and systematically removing nodes from the network, starting with the highest scores. 
Notably, in \cite{holme2002attack}, it was demonstrated that, for synthetic networks as Erd\H{o}s–R\'enyi, Barab\'asi–Albert, or Watts-Strogatz, an attack based on betweenness score generally proves more effective in degrading network performance.

In this section, we replicate a vulnerability analysis by comparing the degree and betweenness strategies with the fitness strategy, considering both the original metric defined in \Eq{\ref{eq:compact_nhefc}} as well as the orthofitness presented in Section \ref{sec:ortho}.
We generated $100$ different artificial networks, each with $200$ nodes. 
This analysis considers random graphs  \cite{erdHos1960evolution}, scale-free \cite{barabasi1999emergence}, and small-world networks \cite{watts1998collective}; the specific parameters used for the simulations are reported in \ref{app:attack}. 
We evaluate network performance under attack using two metrics from \cite{holme2002attack}: the size of the largest connected component $S_c$ and the inverse geodesic distance $\langle l^{-1} \rangle$, representing network navigability and functionality. 
Additionally, we introduce a novel vulnerability measure: the number of connected components in the network $H_c$, which reflects the grade of network disintegration. We also consider the Shannon entropy of the sizes of the connected components $P_c$, which reflects the degree of disruption of the network into components of comparable dimensions. Although ECI and Eigenvector Centrality were employed in the analysis, their results were omitted due to way lower performance. 
Results are shown in Fig. \ref{fig:attack_synthetic}. 
The performance is evaluated with the Area Under the Curve (AUC), and we report the quantitative results in \ref{app:attack}. 

The fitness strategy consistently outperforms the other two in terms of $H_c$. 
In the case of Erd\H{o}s–R\'enyi networks, all three strategies perform similarly, with a slight advantage observed with the fitness. 
In this case, the equivalence of strategies is somewhat expected, given the uniform and compact character of Erd\H{o}s–R\'enyi graphs. 
For Barab\'asi–Albert networks, in $S_c$, $\langle l^{-1} \rangle$ and $P_c$, the fitness strategy generally underperforms compared to betweenness, which proves to be the most effective in these cases. However, it is worth noting that the fitness performs better (or equally) until the removal of $0.3/0.4$ of the nodes. 
In small-world Watts-Strogatz networks, the betweenness strategy excels in terms of $S_c$, $\langle l^{-1} \rangle$ and $P_c$, while fitness outperforms in $H_c$. 

In Table \ref{tab:attack_AUC}, we observe that the fitness strategy demonstrates superior performance compared to the others when metrics are not recomputed at each node removal. 
In this case, the attack utilizes the ranking computed on the entire network at the outset. 
This attack variant proves to be significantly less demanding in terms of computational resources, which can be of considerable interest in specific scenarios. 
Additionally, the time scaling of the fitness and betweenness strategies is depicted in Fig.~\ref{fig:time_scaling}. 
Notably, computing fitness is more efficient than computing betweenness; the resource scaling of the latter is $O(NM)$ where $N$ is the number of nodes and $M$ the number of edges \cite{brandes2001faster}, while the former is observed to scale as $O(N^{0.6}M)$.
Consequently, opting for the fitness strategy may be preferable for large networks due to its ability to yield faster results, although with a slight decrease in performance.
We provide further details on the attack vulnerability analysis in \ref{app:attack}.

%%%%%%%%%%%%%%%%%%%%%%%%%%%%%%%%%%%%%%%%%%
\section{Discussion}
\label{sec:discussion}
\noindent
Estimating the latent capabilities of actors within a network poses considerable challenges. 
These challenges become even more daunting when we rely solely on the network's topological structure, with limited additional information available. 
The Economic Complexity Index (ECI) and the Economic Fitness Complexity (EFC) have proven instrumental in addressing this gap. 
They have been applied to bipartite networks of countries and exported products to indirectly infer nations' inner capabilities. 
Remarkably, despite economists typically relying on many indicators, the scant information on exported products can yield impressive results in economic impact forecasting when approached using the ECI and EFC algorithms.
Before this study, the applicability of ECI and EFC was limited to bipartite networks. 
We have generalized both algorithms to extend their applicability to any graph and utilized the well-known Zachary Karate Club network for testing. 
The Zachary Karate Club network illustrates the social connections among club members, who formed two distinct clusters following an internal dispute. 
Given our knowledge of the resulting split regarding the participants in the new clusters, this network has been widely used as a testbed for community detection algorithms.
The resultant generalized algorithms are outlined in Eqs.~(\ref{eq:eci_transition}) and (\ref{eq:compact_nhefc}). 
In both instances, we observe a loss in interpreting the Product Complexity Index and Economic Complexity of products, respectively, as we now only require one vector. 
The dual sets of quantities - ECI and PCI and Economic Fitness and Economic Complexity - were artifacts arising from the system's bipartite nature.

The generalized ECI is closely linked to the graph's community structure, effectively serving as a spectral-based community detection algorithm. 
Additionally, it can be viewed as a one-dimensional graph embedding and provides the minimum cut of the graph. 
The minimum cut determines the two clusters in the Zachary Karate Club network (refer to Fig. \ref{fig:zkc_eci}, left panel).
ECI functions as a random walk on graphs, failing to fully capture the significance of high-degree nodes. 
For instance, compared to eigenvector centrality — which prioritizes high-degree nodes and their neighbors (refer to the top panel of Fig.~\ref{fig:eigvcentrality}) — ECI assigns more importance to node 30 over node 34 and treats node 1 and 34 differently, despite both having a high number of connections. 
Indeed, nodes 1 and 34 exhibit significantly different rankings according to ECI despite their roles being nearly equivalent in importance. 
After all, nodes 1 and 34 are known to be responsible for the split.
It is worth noting, and we demonstrate in \ref{app:GS} in Eq.~(\ref{eq:orthoECI}), that the ECI vector is orthogonal to the vector of node degrees.

In Fig.~\ref{fig:zkc_fitness}, we illustrate the nodes of the Zachary Karate Club network, with a color code representing their fitness (left panel) and orthofitness (right panel). 
As a fitness or orthofitness vector can be uniquely derived from the network's topology, we refer to them as Fitness Centrality and Orthofitness Centrality, respectively -- introducing two novel centrality measures in graphs. 
It is worth noting how both measures assign greater importance to pivotal nodes 33, 34, and 1.
It comes as no surprise that high-degree nodes hold the utmost importance in the network. 
This observation has led to criticism of EFC, suggesting it does not offer additional valuable information beyond node degree. 
As depicted in the upper left panel of Fig.~\ref{fig:zkc_all_vs_fitness}, Fitness Centrality and node degree exhibit a correlation at high node degrees (fitness greater than 100), whereas the correlation is practically absent at lower degrees (fitness below 10). 
This confirms that fitness is indeed a distinct measure. 
In Fig.~\ref{fig:zkc_all_vs_fitness}, we plot ECI, eigenvector centrality, and orthofitness against fitness. 
In all cases, except for ECI, which represents a fundamentally different measure, the correlation is observed at fitness values greater than 100.
Responding to the critique regarding the non-orthogonality of the fitness vector to the degree vector, we introduced a new metric called Orthofitness Centrality. 
While ECI is automatically orthogonal to the node degree vector, we derive the Orthofitness Centrality by orthogonalizing the fitness vector to the degree vector, a process carried out after the fitness map outlined in \Eq{\ref{eq:compact_nhefc}} has converged. 
It is important to note that this orthogonalization is not performed at every step but only at the end of the map iterations, as orthofitness must encompass components of opposite signs to be orthogonal to a vector with entirely positive entries:
The map of \Eq{\ref{eq:compact_nhefc}} needs positive quantities $\vec{V}$ to work.
In Fig.~\ref{fig:zkc_all_vs_fitness}, we depict ECI, eigenvector centrality, and fitness plotted against orthofitness. 
The correlation is evident for all metrics except ECI at fitness values exceeding 100. 
The plots in the lower right corner of Fig.~\ref{fig:zkc_all_vs_fitness} and Fig.~\ref{fig:zkc_all_vs_orthofitness} are identical, with the axes interchanged, and with fitness plotted in both logarithmic and linear scales, respectively. 
The plot at the lower right in Fig.~\ref{fig:zkc_all_vs_fitness} clarifies that the correlation between fitness and orthofitness occurs at high respective values.
The orthofitness appears to offer little additional insight compared to plain fitness. 
While it mitigates the influence of nodes' degrees on fitness, it does not significantly alter the overall perspective. 

\begin{figure}
    \centering
    \includegraphics[width=8cm]{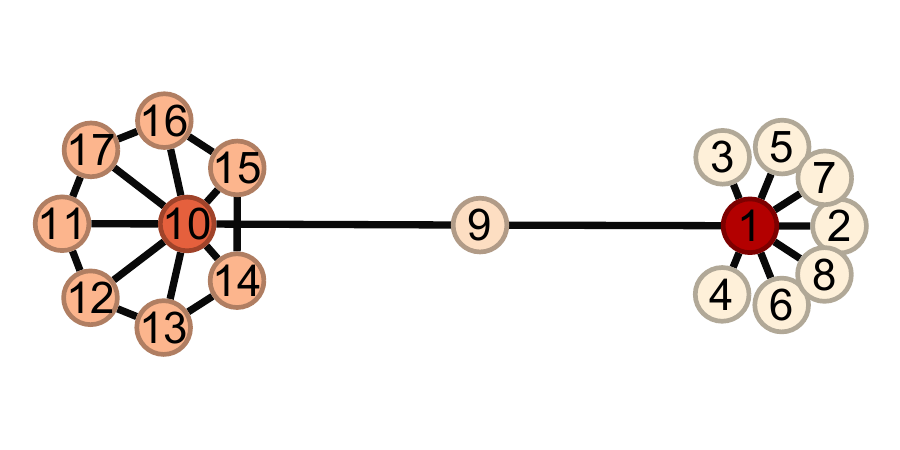}
    \caption{\textbf{Fitness Centrality in a toy graph.}
    The undirected toy network depicted here consists of a wheel and a star connected by a single node. 
    Both nodes 1 and 10 have a degree of 8. 
    Since nodes with lower degrees generally have lower fitness values, and considering that fitness is calculated based on the sum of the reciprocals of a node's neighbors' fitness values, node 1 acquires a higher fitness than node 10.
    Higher fitness centrality is depicted with a darker color.
    }
        \label{fig:toy}
\end{figure}

A natural question arises: Can fitness centrality provide additional valuable information beyond node degree?
To further clarify this point, we computed the Fitness Centrality in a toy graph comprising a star connected to a wheel (see Fig.~\ref{fig:toy}).
Node 1 at the center of the star obtains a higher fitness compared to node 10 at the center of the wheel because it is linked to nodes with lower degrees, which in turn have lower fitness values. 
We recall that a node's fitness centrality is calculated as the sum of the reciprocals of its neighbors' fitness values.
Essentially, high-fitness nodes are more likely to disrupt the network if removed. 
In our toy graph, removing node 1 detaches seven nodes from the network, whereas removing node 10 detaches the wheel from the network, but its neighbors remain connected in a circle.
We investigated the ability of high-fitness nodes to disrupt the network when removed using three synthetic networks, and the results are presented in Figure \ref{fig:attack_synthetic}. 
Indeed, fitness-based node removal outperforms degree-based node removal in network disruption and shows comparable performance to node betweenness. 
Fitness surpasses node betweenness when the quantities are not recalculated after each node removal, as demonstrated in Table~\ref{tab:attack_AUC}. 
This method is computationally less demanding and could benefit large networks where the computation of node betweenness is slow.
High-fitness node fitness can be estimated accurately after just a few iterations of the map in \Eq{\ref{eq:compact_nhefc}}. 
At the same time, longer computational times are needed to achieve convergence for low-fitness nodes, which have negligible importance for network disruption.

It is interesting to take a step back and interpret the cost functions of ECI and NHEFC, as described in Eqs.~(\ref{eq:eci_cost_general}) and (\ref{eq:efc_cost_general}), specifically in the context of the bipartite network involving countries and exported products. 
We can reformulate the ECI cost function from Eq.~(\ref{eq:eci_cost_general}) in terms of the economic complexity index $F_c$ for countries and the product complexity index $Q_p$ for the products they export:
\begin{equation}
    U(\vec{F}, \vec{Q}) = - \dfrac{1}{2} \sum_{c, p} M_{cp} F_c Q_p + \dfrac{1}{2}\sum_c k_c F_c^2 + \dfrac{1}{2}\sum_p k_p Q_p^2
    \label{eq:eci_cost}
\end{equation}
The first sum represents a conventional ferromagnetic coupling between ECI and PCI. In line with the linear map described in Equation (\ref{eq:eci}), it tends to align ECI and PCI, meaning that large ECI values lead to large PCI values.

The case of NHEFC is more interesting.
The of \Eq{\ref{eq:efc_cost_general}} for the country-product bipartite network is:
\begin{equation}
    U(\vec{F}, \vec{Q}) = \dfrac{1}{2} \sum_{c, p} M_{cp} \dfrac{Q_p}{F_c} - \sum_c \log \dfrac{1}{F_c} - \sum_p \log Q_p + \delta \sum_c \dfrac{1}{F_c} + \delta \sum_p Q_p
    \label{eq:efc_cost}
\end{equation}
The connection between EFC algorithms and optimization problems has been shown by exploiting the similarity with the Sinkhorn-Knopp algorithm for matrix rescaling \cite{knight2008sinkhorn}. 
In \cite{mazzilli2024equivalence}, the authors gave an interesting economic interpretation of the interactions $\sum_{c,p} M_{cp} Q_p / F_c$; each term $Q_p/F_c$ represents the energy cost for country $c$ to export product $p$. 
The minimization of the cost function requires countries with low fitness not to be linked to high-complexity products since they do not have the capabilities to produce them. 
Each country distributes its capabilities most efficiently. 
This is related to the EFC algorithm's ability to reorder a matrix's rows and columns to enhance its triangular structure. 
This feature can be explained in terms of Kantorovich potentials \cite{staudt2022uniqueness}.

%%%%%%%%%%%%%%%%%%%%%%%%%%%%%%%%%%%%%%%%%%
\section{Conclusions}
% This section is not mandatory, but can be added to the manuscript if the discussion is unusually long or complex.
\noindent
In this study, we expand the scope of economic complexity algorithms to encompass general graphs, whereas they were previously applicable only to bipartite graphs. 
We formulate the Economic Complexity Index (ECI) and Economic Fitness Complexity (EFC) in a generalized form involving the graph's adjacency matrix. 
This generalization extends the significance of Fitness Complexity, transforming it into a novel centrality measure in graphs that we call Fitness Centrality. 
At the same time, the extended ECI remains a one-dimensional embedding of graphs. 
Inspired by ECI being orthogonal to the node degree vector, we introduce a novel economic complexity metric termed Orthofitness Centrality, derived by orthogonalizing the fitness vector relative to the node degree vector. 
However, we do not find significant benefits of orthofitness compared to fitness. 
Conversely, Fitness Centrality can effectively assess network vulnerability, nearly matching the performance of node betweenness centrality in detecting the most influential nodes. 
We also identify cost functions minimized during the evolution of both ECI and NHEFC recursive maps. 
Notably, in the ECI algorithm, the product of ECI times PCI tends to be large. 
In contrast, the ratio between the complexity of exported products and the fitness of exporting countries tends to be small in the EFC algorithm. 
Essentially, countries with small fitness are associated with products with small complexity.

For further analysis in the future, the structure of the generalized Fitness map allows for straightforward modification. 
For example, the nonlinear term can be adjusted as $(\vec{V})^{-\nu}$, where $0 < \nu < 1$ diminishes the negative significance of low-degree nodes, while $\nu > 1$ amplifies their importance.
Another possible extension is to apply it to directed and weighted graphs.
Also, it might be insightful to test the effectiveness of fitness-based node removal in networks with very different topologies.

%%%%%%%%%%%%%%%%%%%%%%%%%%%%%%%%%%%%%%%%%%
\vspace{6pt} 

%%%%%%%%%%%%%%%%%%%%%%%%%%%%%%%%%%%%%%%%%%
%% optional
%\supplementary{The following supporting information can be downloaded at:  \linksupplementary{s1}, Figure S1: title; Table S1: title; Video S1: title.}

%%%%%%%%%%%%%%%%%%%%%%%%%%%%%%%%%%%%%%%%%%
\paragraph{Author contributions:}
Conceptualization, V.D.P.S.; 
methodology, V.D.P.S. and A.B.; 
software, V.D.P.S., E.C. and A.B.; 
validation, V.D.P.S., E.C. and G.D.M.; 
formal analysis, V.D.P.S. and A.B.; 
investigation, E.C. and A.B.; 
resources, G.D.M.; 
data curation, V.D.P.S. and A.B.; 
writing---original draft preparation, E.C.; 
writing---review and editing, V.D.P.S., G.D.M. and A.B.; 
visualization, V.D.P.S. and A.B.; 
supervision, V.D.P.S.; 
project administration, V.D.P.S.; 
funding acquisition, V.D.P.S. and E.C.;
All authors have read and agreed to the published version of the manuscript.

\paragraph{Funding:}
This work was funded by the City of Vienna, Municipal Department 7, and the Federal Ministry of the Republic of Austria for Climate Action, Environment, Energy, Mobility, Innovation and Technology as part of the project GZ~2021-0.664.668.

\paragraph{Acknowledgments:}
We acknowledge valuable discussions with C.~Chilin, D.~Mazzilli, and A.~Patelli.

\paragraph{Conflicts of interest:}
The authors declare no conflict of interest.

\paragraph{Abbreviations:}
The following abbreviations are used in this manuscript:\\

\noindent 
\begin{tabular}{@{}ll}
    ECI & Economic Complexity Index\\
    EFC & Economic Fitness Complexity\\
    NHEFC & Non-Homogeneous Economic Fitness Complexity\\
    PCI & Product Complexity Index
    % RCA & Revealed Comparative Advantage
\end{tabular}
%%%%%%%%%%%%%%%%%%%%%%%%%%%%%%%%%%%%%%%%%%

\bibliographystyle{unsrt}
% \bibliography{main}  

\renewcommand\thesection{Appendix~\Alph{section}}
\setcounter{section}{0}

\section{The Zachary Karate Club dataset}
\label{app:data}
\noindent
We employed the widely recognized Zachary karate club network, sourced from data gathered by Wayne Zachary in 1977 from the members of a university karate club \cite{konect2017zachary,zachary1977information}. 
In this network, each node stands for a club member, and each edge signifies a connection between two members. 
The network is undirected. 
This dataset is often used as a test-bed for community detection algorithms. 
It involves identifying the two factions that emerged within the karate club following a dispute between two instructors.

%%%%%%%%%%%%%%%%%%%%%%%%%%%%%%%%%%%%%%%%%%

\section{Courant–Fischer–Weyl process to find eigenvectors of a normal matrix}
\label{app:GS}
\noindent
In the following, we will define a straightforward way to estimate the first eigenvectors of normal matrices without solving for the whole spectrum.
We show this procedure here since normal matrices are not symmetric.
We focus on $\vec{F}^{(n)}$, that is, on countries (analogous results hold for the products).
We start by noticing that the eigenvectors of $\mathbf{N}_1$ are not orthogonal since $\mathbf{N}_1$ is not symmetric, we thus define the symmetric matrix $\mathbf{H}$ 
\begin{equation}
    \mathbf{H}=\mathbf{D}^{-\frac{1}{2}}\mathbf{S}_1\mathbf{D}^{-\frac{1}{2}}=\mathbf{D}^{+\frac{1}{2}}\mathbf{N}_1\mathbf{D}^{-\frac{1}{2}}.
    \label{eq:hN}
\end{equation}
The matrix $\mathbf{H}$ is symmetric, thus its eigenvectors are orthogonal. 
Moreover, it has the same eigenvalues of $\mathbf{N}_1$. 
Indeed, the eigenvalue problem for $\mathbf{H}$ can be expressed as
\begin{equation}
    \mathbf{H}\vec{\psi}_i=\lambda_i\vec{\psi}_i,
\end{equation}
which, using Eq. \eqref{eq:hN}, becomes
\begin{equation}
    \mathbf{D}^{\frac{1}{2}}\mathbf{N}_1\mathbf{D}^{-\frac{1}{2}}\vec{\psi}_i=\lambda_i\vec{\psi}_i.
\end{equation}
Thus
\begin{equation}
    \mathbf{N}_1\bigl(\mathbf{D}^{-\frac{1}{2}}\vec{\psi}_i\bigr)=\lambda_i\bigl(\mathbf{D}^{-\frac{1}{2}}\vec{\psi}_i\bigr) \qquad \to \qquad \mathbf{N}_1\vec{\varphi}_i =\lambda_i \vec{\varphi}_i,
    \label{eq:normal_eigenvect}
\end{equation}
where we defined $\vec{\varphi}_i = \mathbf{D}^{-\frac{1}{2}}\vec{\psi}_i$. 

Exploiting the orthogonality of $\vec{\psi}_i$, we have
\begin{equation}
    \delta_{ij} = \bigl(\vec{\psi}_i\bigr)^T\vec{\psi}_j = \bigl(\vec{\varphi}_i\bigr)^T\mathbf{D}^{\frac{1}{2}}\mathbf{D}^{\frac{1}{2}}\vec{\varphi}_j = \bigl(\vec{\varphi}_i\bigr)^T\mathbf{D}\vec{\varphi}_j,
\end{equation}
implying 
\begin{equation}
    \bigl(\vec{\varphi}_i\bigr)^T\mathbf{D}\vec{\varphi}_j = \delta_{ij}.
    \label{eq:orthogonal}
\end{equation}
Thus, the eigenvectors of $\mathbf{N}_1$, $\vec{\varphi}_i$, can also be orthogonalized if appropriately multiplied by $\mathbf{D}$, which plays the role similar to a metric tensor. 
To determine the second eigenvector, we start from a random vector $\vec{r}^{(0)}$ and iterate the following map involving the vector $\vec{d}$ of the elements in the diagonal of $\mathbf{D}$:
\begin{equation}
    \vec{s}^{(n)} = \vec{r}^{(n)} - \Biggl(\frac{\vec{r}^{(n)}\cdot\vec{d}}{\bigl|\vec{d}\bigr|^2} \Biggr)\vec{d},
    ~~~
    \vec{t}^{(n)} = \frac{\vec{s}^{(n)} } {|\vec{s}^{(n)} |},
    ~~~
    \vec{r}^{(n+1)} = \mathbf{N}_1 \vec{t}^{(n)}.
\end{equation}
In the previous procedure, we first remove the components of the constant eigenvector with a Gram-Schmidt process. We then normalize the result with a $L_2$ norm and finally apply the matrix.
We can generalize the procedure to find all other eigenvectors by orthogonalizing the vector $\vec{r}$ with respect to the known ones.
If needed, the value of the second eigenvalue can be estimated at the end of the previous procedure by using the following limit of the Rayleigh-Ritz quotient
\begin{equation}
    \lambda_2 = \lim_{n\to\infty} \frac{|\mathbf{N}_1 \vec{r}^{(n)}|}{|\vec{r}^{(n)}|}.
\end{equation}
As a direct consequence of \Eq{\ref{eq:orthogonal}}, we observe that the diversification vector is orthogonal to the ECI. 
Specifically, \Eq{\ref{eq:orthogonal}} between the first (constant and normalized) eigenvector and the second (ECI) can be expressed as:
\begin{equation}
    \bigl(\vec{\varphi}_1\bigr)^T\mathbf{D}\vec{\varphi}_2 = 
    c \bigl(\vec{d}\,\bigr)^T  \vec{\varphi}_2 = 0,
    \label{eq:orthoECI}
\end{equation}
with a $c$ as a multiplicative constant.
All we showed for the matrix $\mathbf{N}_1$ and ECI applies also to $\mathbf{N}_2$ and PCI.

%%%%%%%%%%%%%%%%%%%%%%%%%%%%%%%%%%%%%%%%%%

\section{Derivation of Cost Functions}
\label{app:cost_derivation}
\noindent
In this appendix, we provide the analytical details of the computations for both ECI and NHEFC algorithms. We start with the simplest case, namely ECI.

To express the cost function, we must formulate \Eq{\ref{eq:eci_map}} as a map. This involves subtracting $V_i^{(n-1)}$ from both sides of the equations and considering the left-hand side $V_i^{(n)} - V_i^{(n-1)} \approx \frac{d}{dn}V_i = \dot{V_i}^{(n-1)}$ as an approximate derivative. The resulting map is expressed as:
\begin{equation}
\dot{V_i}^{(n-1)} = \dfrac{1}{k_i} \sum_{j=1}^N A_{ij} V_j^{(n-1)} - V_i^{(n-1)}
\label{eq:eci_field}
\end{equation}
Here, the dot represents the derivative with respect to the iteration index $^{(n)}$, and $N = N_c + N_p$ denotes the total number of nodes in the network, irrespective of their specific nature. The right-hand side of the equation can be interpreted as the vector field driving the dynamics to convergence.

It is convenient to introduce a change of variable to make the adjacency matrix of the network symmetric. 
This is required in order to compute a cost function of the map. 
Introducing the variables $Z_i = \sqrt{k_i} V_i$, \Eq{\ref{eq:eci_field}} reads as:
\begin{equation}
\dot{Z}_i = \sum_{j=1}^N \dfrac{A_{ij}}{\sqrt{k_i k_j}} Z_j - Z_i
\label{eq:eci_field_2}
\end{equation}
The potential can now be obtained by integrating this map, but the Schwartz condition must be satisfied. 
This is equivalent to require that the vector field is irrotational (to be more precise, the 2-form derived from the vector field 1-form is zero):
\begin{equation}
\dfrac{\partial \dot{Z_i}}{\partial Z_h} = \dfrac{\partial \dot{Z_h}}{\partial Z_i}
\label{eq:schwarz}
\end{equation}
Given that the full adjacency matrix $A_{ij}$ is symmetric by construction, both quantities are equal to $\dfrac{A_{ih}}{\sqrt{k_i k_h}} + \delta_{ih}$. 
This ensures the existence of a potential function in the variables $Z_i$ with the property that $- \nabla_{Z_i} U = \dot{Z_i}$. 
From this we recover the explicit expression by integrating \Eq{\ref{eq:eci_field_2}}:
\begin{equation}
U(\vec{Z}) = - \dfrac{1}{2} \sum_{i, j} \dfrac{A_{ij}}{\sqrt{k_i k_j}} Z_i Z_j + \dfrac{1}{2}\sum_i Z_i^2.
\label{eq:ECI_potential_Z}
\end{equation}
The factor $1/2$ avoids counting each link two times since the matrix $A_{ij}$ is symmetric. 
Replacing back the original variables $V_i = Z_i / \sqrt{k_i}$, the cost functions is expressed as:
\begin{equation}
U(\vec{V}) = - \dfrac{1}{2} \sum_{i, j} A_{ij} V_i V_j + \dfrac{1}{2}\sum_i k_i V_i^2
\label{eq:ECI_potential_V}
\end{equation}
We remark that the interpretation as a potential function is possible only using the natural variables $Z_i$. 
The minimum of \Eq{\ref{eq:ECI_potential_Z}} coincides with the minimum of \Eq{\ref{eq:ECI_potential_Z}}, but the dynamics leading to the minimization may not be equivalent.

For NHEFC, the map equivalent to \Eq{\ref{eq:eci_field}} is:
\begin{equation}
\dot{V_i} = \delta + \sum_{j=1}^N A_{ij} V_j^{-1} - V_i
\label{eq:efc_field}
\end{equation}
The vector field associated with the system is not irrotational, as \Eq{\ref{eq:schwarz}} are not satisfied. Through an appropriate change of variable $Z_i = - \log V_i$, \Eq{\ref{eq:efc_field}} can be transformed as
\begin{equation}
\dot{Z_i} = 1 - \delta e^{Z_i} - e^{Z_i}\sum_{j=1}^N A_{ij} e^{Z_j}
\label{eq:efc_field_z}
\end{equation}
This map is now irrotational, and the change of variable naturally reveals the intrinsic logarithmic scale of EFC. 
Integrating this expression gives:
\begin{equation}
U(\vec{Z}) = \dfrac{1}{2} \sum_{i, j} A_{ij} e^{Z_i} e^{Z_j} - \sum_i Z_i + \delta \sum_i e^{Z_i}
\end{equation}
Finally, \Eq{\ref{eq:efc_cost_general}} can be obtained by replacing the original variables $V_i = e^{-Z_i}$.

%%%%%%%%%%%%%%%%%%%%%%%%%%%%%%%%%%%%%%%%%%

\section{Comparison with eigenvector centrality}
\label{app:eigvc}

\begin{figure}
    \centering
    \includegraphics[width=0.6\textwidth]{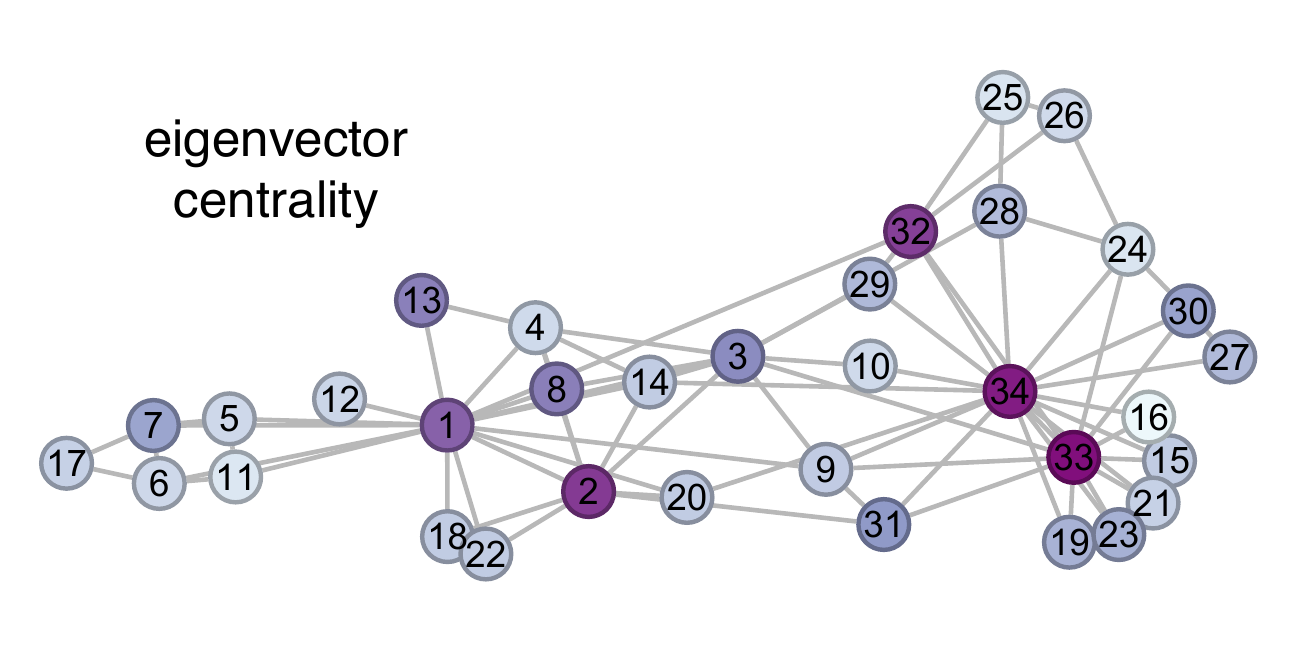}\\
    \centering
    \includegraphics[width=0.9\textwidth]{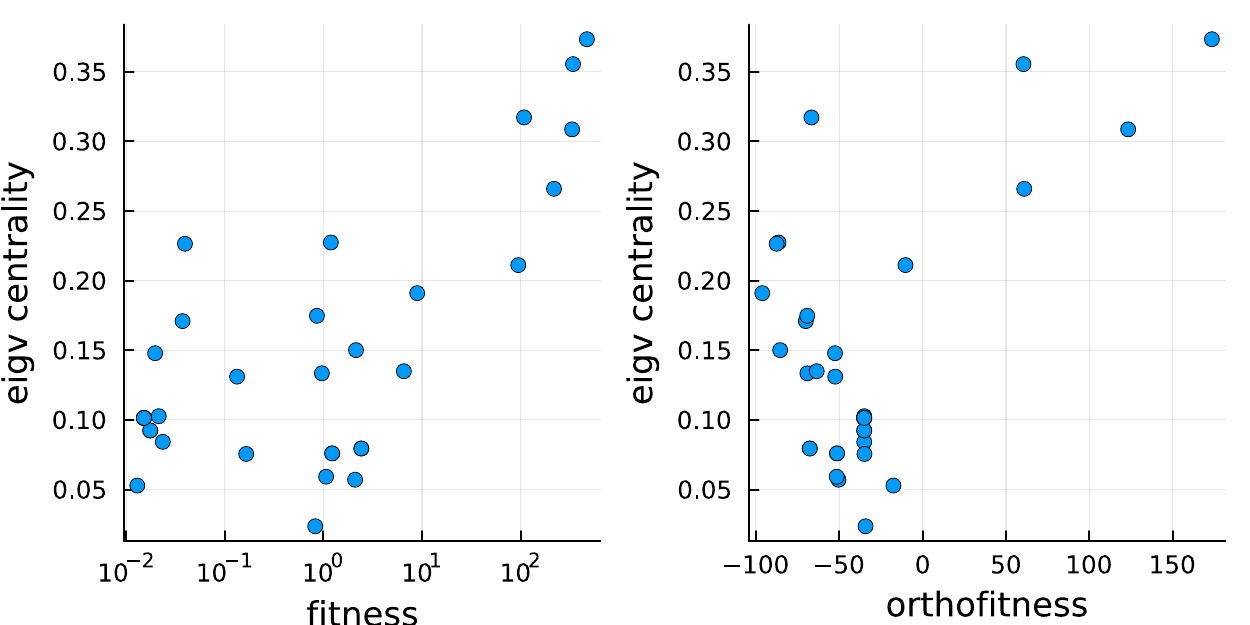}
    \caption{\textbf{Additional comparison between fitness and eigenvector centrality}
    In the top panel, the Zachary Karate Club graph nodes were color-coded based on their eigenvector centrality scores, derived from the principal eigenvector of the adjacency matrix. 
    Darker shades indicate higher centrality scores. 
    Eigenvector centrality tends to prioritize nodes with higher degrees and their neighboring nodes. 
    In the lower panels, fitness scores correlate with eigenvector centrality at higher values.
    Compared to eigenvector centrality, both fitness scores assign relatively less significance to the neighbors of high-degree nodes.
    }
    \label{fig:eigvcentrality}
\end{figure}
\noindent
In this section, we show in Fig.~\ref{fig:eigvcentrality} the comparison between fitness and orthofitness against the eigenvector centrality.

%%%%%%%%%%%%%%%%%%%%%%%%%%%%%%%%%%%%%%%%%%

\section{Attack Vulnerability of networks}
\label{app:attack}
\noindent

\begin{table}[t!]
\begin{tabularx}{\textwidth}{|>{\centering\arraybackslash}X|>{\centering\arraybackslash}X|>{\centering\arraybackslash}X|>{\centering\arraybackslash}X|>{\centering\arraybackslash}X|}
    
    \multicolumn{5}{c}{\textbf{Area Under Curve (AUC) (R attack) }} \\
    \hline
    \textbf{Erd\H{o}s–R\'enyi} & $\boldsymbol{H_c}$ & $\boldsymbol{S_c}$ & $\boldsymbol{\langle l^{-1} \rangle}$ & $\boldsymbol{P_c}$\\
    \hline
    Degree & $0.517 \pm 0.001$ & $0.521 \pm 0.001$ & $0.582 \pm 0.005$ & $0.621 \pm 0.001$ \\
    Betweenness & $0.518 \pm 0.002$ & $0.525 \pm 0.002$ & $0.586 \pm 0.005$ & $0.624 \pm 0.001$\\
    Fitness  & $0.526 \pm 0.002$ & $0.526 \pm 0.002$ & $\boldsymbol{0.588 \pm 0.005}$ & $0.626 \pm 0.001$\\
    OrthoFitness & $\boldsymbol{0.535 \pm 0.002}$ & $\boldsymbol{0.535 \pm 0.002}$ & $0.582 \pm 0.007$ & $\boldsymbol{0.634 \pm 0.001}$\\
    \hline
    \textbf{Barab\'asi–Albert} & $\boldsymbol{H_c}$ & $\boldsymbol{S_c}$ & $\boldsymbol{\langle l^{-1} \rangle}$ & $\boldsymbol{P_c}$\\
    \hline
    Degree & $0.668 \pm 0.009$ & $0.711 \pm 0.009$ & $0.906 \pm 0.004$ & $0.753 \pm 0.007$\\
    Betweenness & $0.650 \pm 0.008$ & $\boldsymbol{0.726 \pm 0.008}$ & $\boldsymbol{0.906 \pm 0.004}$ & $\boldsymbol{0.760 \pm 0.007}$\\
    Fitness & $\boldsymbol{0.686 \pm 0.007}$ & $0.691 \pm 0.008$ & $0.891 \pm 0.005$ & $0.752 \pm 0.006$\\
    OrthoFitness & $0.676 \pm 0.008$ & $0.686 \pm 0.009$ & $0.872 \pm 0.007$ & $0.751 \pm 0.006$\\
    \hline
    \textbf{Watts-Strogatz} & $\boldsymbol{H_c}$ & $\boldsymbol{S_c}$ & $\boldsymbol{\langle l^{-1} \rangle}$ & $\boldsymbol{P_c}$\\
    \hline
    Degree & $0.598 \pm 0.004$ & $0.670 \pm 0.012$ & $0.921 \pm 0.004$ & $0.715 \pm 0.007$\\
    Betweenness & $0.612 \pm 0.004$ & $\boldsymbol{0.832 \pm 0.013}$ & $\boldsymbol{0.948 \pm 0.004}$ & $\boldsymbol{0.822 \pm 0.010}$\\
    Fitness & $0.639 \pm 0.005$ & $0.674 \pm 0.013$ & $0.914 \pm 0.005$ & $0.732 \pm 0.007$\\
    OrthoFitness & $\boldsymbol{0.641 \pm 0.004}$ & $0.648 \pm 0.009$ & $0.889 \pm 0.009$ & $0.723 \pm 0.005$\\
    \hline
        
\end{tabularx}

\vspace{0.5cm}

\begin{tabularx}{\textwidth}{|>{\centering\arraybackslash}X|>{\centering\arraybackslash}X|>{\centering\arraybackslash}X|>{\centering\arraybackslash}X|>{\centering\arraybackslash}X|}
    
    \multicolumn{5}{c}{\textbf{Area Under Curve (AUC) (I attack) }} \\
    \hline
    \textbf{Erd\H{o}s–R\'enyi} & $\boldsymbol{H_c}$ & $\boldsymbol{S_c}$ & $\boldsymbol{\langle l^{-1} \rangle}$ & $\boldsymbol{P_c}$\\
    \hline
    Degree & $0.507 \pm 0.001$ & $0.508 \pm 0.001$ & $0.516 \pm 0.009$ & $0.615 \pm 0.001$ \\
    Betweenness & $0.508 \pm 0.001$ & $0.509 \pm 0.001$ & $\boldsymbol{0.532 \pm 0.009}$ & $0.616 \pm 0.001$\\
    Fitness  & $\boldsymbol{0.508 \pm 0.001}$ & $\boldsymbol{0.509 \pm 0.001}$ & $0.530 \pm 0.010$ & $\boldsymbol{0.616 \pm 0.001}$\\
    OrthoFitness & $0.507 \pm 0.001$ & $0.508 \pm 0.001$ & $0.496 \pm 0.015$ & $0.615 \pm 0.001$\\
    \hline
    \textbf{Barab\'asi–Albert} & $\boldsymbol{H_c}$ & $\boldsymbol{S_c}$ & $\boldsymbol{\langle l^{-1} \rangle}$ & $\boldsymbol{P_c}$\\
    \hline
    Degree & $0.643 \pm 0.011$ & $0.686 \pm 0.011$ & $0.897 \pm 0.005$ & $0.737 \pm 0.008$\\
    Betweenness & $0.641 \pm 0.011$ & $0.686 \pm 0.011$ & $0.897 \pm 0.005$ & $0.736 \pm 0.008$\\
    Fitness & $\boldsymbol{0.670 \pm 0.010}$ & $\boldsymbol{0.698 \pm 0.009}$ & $\boldsymbol{0.901 \pm 0.004}$ & $\boldsymbol{0.749 \pm 0.007}$\\
    OrthoFitness & $0.596 \pm 0.015$ & $0.600 \pm 0.018$ & $0.803 \pm 0.030$ & $0.691 \pm 0.012$\\
    \hline
    \textbf{Watts-Strogatz} & $\boldsymbol{H_c}$ & $\boldsymbol{S_c}$ & $\boldsymbol{\langle l^{-1} \rangle}$ & $\boldsymbol{P_c}$\\
    \hline
    Degree & $0.559 \pm 0.005$ & $0.657 \pm 0.022$ & $0.910 \pm 0.007$ & $0.704 \pm 0.012$\\
    Betweenness & $0.543 \pm 0.004$ & $0.668 \pm 0.018$ & $0.901 \pm 0.005$ & $0.703 \pm 0.010$\\
    Fitness & $\boldsymbol{0.590 \pm 0.006}$ & $\boldsymbol{0.683 \pm 0.018}$ & $\boldsymbol{0.922 \pm 0.005}$ & $\boldsymbol{0.725 \pm 0.011}$\\
    OrthoFitness & $0.589 \pm 0.007$ & $0.638 \pm 0.017$ & $0.896 \pm 0.008$ & $0.702 \pm 0.010$\\
    \hline
        
\end{tabularx}

    \vspace{0.3cm}
    
    \caption{\textbf{AUC performances of degree, betweenness and fitness attack strategies.} In this table are reported the Area Under the Curve values obtained from the attack vulnerability analysis on three different types of networks. In each case, we consider a sample of $100$ random networks. The upper table reports the result for the "recalculated" (R) attack, where the metrics are recomputed at each node removal. The fitness generally performs better in terms of $H_c$, while for $S_c$, $\langle l^{-1}\rangle$ and $P_c$, it is more effective in the initial stage of the attack. The lower table reports the result for the "initial" (I) attack, where we consider the metrics computed on the whole network throughout all the attacks. The fitness generally outperforms any other strategies for $H_c$, $S_c$, $\langle l^{-1}\rangle$ and $P_c$. The best average performance is given in bold text.}
    \label{tab:attack_AUC}
\end{table}
\noindent
In this appendix, we provide additional details regarding the vulnerability analysis on the network discussed in Section \ref{sec:attack}. The attack procedure works as follows: a) a node metric, which can be the degree, betweenness, or fitness, is computed on the network; b) the node with the highest score is removed; c) the process is iterated until all the nodes are disconnected. This is equivalent to the "recalculated" variant (R) presented in \cite{holme2002attack}, which typically performs better than the "initial" variant (I), where the metrics are not recomputed after a node is removed from the network. All the analyses are repeated for both the original fitness \Eq{\ref{eq:compact_nhefc}} and the orthofitness presented in Section \ref{sec:ortho}. Since both methods lead to similar results, we will generally refer to them as ``the fitness strategy''.   

To assess the vulnerability of the graph under attack, we use the size of the largest connected component $S_c$, which reveals how many nodes remain in the largest connected part of the network. We adopt also the inverse geodesic length, which is defined as:
\begin{equation}
    \langle l^{-1} \rangle = \dfrac{1}{N(N-1)} \sum_{i \neq j} \dfrac{1}{d_{ij}}
    \label{eq:inverse_geodesic_length}
\end{equation}
where $d_{ij}$ is the shortest path between node $i$ and $j$. This quantity is the average value of the inverse of the distance between nodes in such a way that two disconnected nodes contribute to $0$. % instead of $\infty$. 
A large value of $\langle l^{-1}\rangle$ reflects high functionality and navigability in the network, while it approaches $0$ when all nodes are disconnected.

We also considered the number of connected components $H_c$ as a measure of the grade of network disruption under attack and the Shannon entropy of the component sizes $P_c$. Given the sizes of each component $V_i$, this is defined as
\begin{equation}
    P_c = - \dfrac{1}{\log H_c} \sum_{i=1}^{H_c} \left( \dfrac{V_i}{N} \right) \log \left( \dfrac{V_i}{N} \right)
\end{equation}
where $N$ is the number of nodes in the network. The sum is calculated for all the $H_c$ connected components. The entropy is normalized to obtain a value between $0$ and $1$. This measure reflects the homogeneity in the sizes of the components that disrupt the network.

Performance is evaluated in Fig.~\ref{fig:attack_synthetic} averaging over $100$ different networks of the same type and computing the Area Under the Curve (AUC). 
This value is $1$ if the network is disrupted at the first step, while it is $0$ if it is utterly resilient to the attack. We report in Table \ref{tab:attack_AUC} the AUC with errors for the cases considered in Fig. \ref{fig:attack_synthetic}. 
In the case of $S_c$ and $\langle l^{-1} \rangle$, we considered $1$ minus the area under the curve as the performance score in such a way that larger AUC values always reflect higher performance (see Fig \ref{fig:attack_synthetic}). 
The errors are computed as the standard deviation on the sample of $100$ networks. When two strategies have statistically comparable performances, they can perform equally on a single attack, but the strategy with the largest average performance is the most efficient \emph{a priori}.

All the networks considered in the analysis contain $N = 200$ nodes. The parameters used to generate the networks are: for the Erd\H{o}s–R\'enyi, $p = 0.1$; for the Barab\'asi–Albert, $m=5$; for the Watts-Strogatz, $p=0.1$, $k = 6$. 
We extended the analysis considering other values of the parameters: $p=0.01, 0.025, 0.5, 0.1$ for Erd\H{o}s–R\'enyi; $m = 3, 4, 5, 6$ for Barab\'asi–Albert; $p=0.01, 0.025, 0.5, 0.1$ for Watts-Strogatz. 
These cases all yield qualitatively equivalent results (the relative effect of the strategies stays the same), therefore we omit their detailed presentation.

\begin{figure}[t!]
    \centering
    \includegraphics[width=0.9\textwidth]{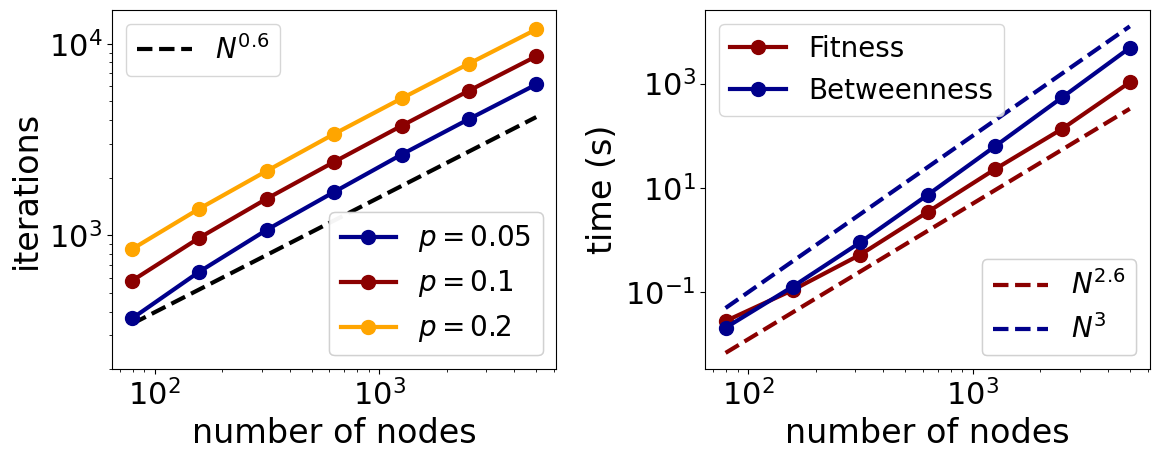}
    \caption{\textbf{Computational resource scaling for fitness and betwenness strategies.}
        \textbf{Left} Number of iterations needed for the fitness algorithms to converge as a function of the number of nodes in the network. The convergence parameters used for this analysis are $\delta = 10^{-2}$, $\epsilon = 10^{-2}$, see \cite{servedio2018new}. The black solid line represents the scaling $N^{0.6}$; different curves represent different values of the parameter $p$ of the random network. \textbf{Right} Time to compute the betweenness and the fitness of every node in the network ($p=0.1$). The Brandes' algorithm for the betweenness scales as $O(N^3)$, while the fitness algorithms only as $O(N^{2.6})$.
        }
    \label{fig:time_scaling}
\end{figure}

The final observation refers to the time scaling of the fitness and betweenness strategies. Brandes' algorithm \cite{brandes2001faster} is the most efficient way to calculate the betweenness of each node, with a computational scaling of $O(NM)$, where $N$ is the number of nodes and $M$ is the number of edges in the network. Conversely, the fitness algorithm performs $O(M)$ operations at each iteration. The number of iterations required for convergence scales as $N^{0.6}$, as illustrated in Fig. \ref{fig:time_scaling}. Consequently, the scaling of the fitness algorithm is $O(N^{0.6}M)$. 

To test the resource scaling of the two algorithms, we generate $7$ Erd\H{o}s–R\'enyi random networks with $p=0.1$ ranging from $80$ to $5000$ nodes. We also repeated the analysis for different values of $p$ to test the universality of the scaling exponent $0.6$. We measure the time required to compute betweenness and fitness in each network. For Erd\H{o}s–R\'enyi networks $M = O(N^2)$, thus the betweenness scales as $O(N^{3})$ while the fitness algorithm only as $O(N^{2.6})$. This could represent a significant speed-up when the number of nodes in the network is large.

%%%%%%%%%%%%%%%%%%%%%%%%%%%%%%%%%%%%%%%%%%

\end{document}